\documentclass[11 pt]{article}
\usepackage[mathscr]{eucal}
\usepackage{amssymb,latexsym}
\usepackage{verbatim}
\usepackage{amsmath}
\usepackage{amsthm}
\usepackage{enumerate}
\usepackage{authblk}
\usepackage{color}
\usepackage{multicol}
\usepackage{tikz}
\usepackage{xypic}

\newtheorem{thm}{Theorem}[section]
\newtheorem{Def}[thm]{Definition}
\newtheorem{prop}[thm]{Proposition}

\renewcommand\S{\Sigma}

\newcommand\s{\sigma}

\newcommand\e{\varepsilon}
\renewcommand\b{\beta}

\newcommand\g{\gamma}

\renewcommand\a{\alpha}

\renewcommand\t{\tau}

\newcommand\beq{\begin{equation}}
\newcommand\eeq{\end{equation}}
\newcommand\ben{\begin{enumerate}}
\newcommand\een{\end{enumerate}}
\newcommand\bit{\begin{itemize}}
\newcommand\eit{\end{itemize}}

\newcommand{\R}{\mathbb R}

\newcommand{\ms}{\mathscr}

\newcommand{\ext}{\text{{\rm ext}}}
\newcommand{\pd}{\partial}

\newcommand{\mc}{\mathcal}
\newcommand{\mf}{\mathfrak}

\newcommand{\Z}{\mathbb{Z}}
\newcommand{\C}{\mathbb{C}}

\def\undertilde#1{\mathord{\vtop{\ialign{##\crcr
   $\hfil\displaystyle{#1}\hfil$\crcr\noalign{\kern1.5pt\nointerlineskip}
   $\hfil\tilde{}\hfil$\crcr\noalign{\kern1.5pt}}}}}

\newcounter{mnotecount}

\title{Milne-like Spacetimes and  their Symmetries}

\author{Eric Ling\footnote{eling@math.miami.edu}}
\affil{Department of Mathematics
\\ University of Miami }

\begin{document}
\date{}
\maketitle
\vspace{.2in}

\begin{abstract} 
When developing a quantum theory for a physical system, one determines the system's symmetry group and its irreducible unitary representations. For Minkowski space, the symmetry group is the Poincar{\'e} group, $\R^4 \rtimes \text{O}(1,3)$, and the irreducible unitary representations are interpreted as elementary particles which determine the particle's mass and spin. We determine the symmetry group for Milne-like spacetimes, a class of cosmological spacetimes, to be $\R \times \text{O}(1,3)$ and classify their irreducible unitary representations. Again they represent particles with mass and spin. Unlike the classification for the Poincar{\'e} group, we do not obtain any faster-than-light particles. The factor $\R$ corresponds to cosmic time translations. These generate a mass Casimir operator which yields a Lorentz invariant Dirac equation on Milne-like spacetimes. In fact it's just the original Dirac equation multiplied by a conformal factor $\Omega$. Therefore many of the invariants and symmetries still hold. We offer a new interpretation of the negative energy states and propose a possible solution to the matter-antimatter asymmetry problem in our universe.

\end{abstract}

\tableofcontents

\section{Introduction}

Milne-like spacetimes are a class of FLRW models which admit continuous spacetime extensions through the big bang. This was first proved in \cite{GalLing_con}. In this paper we explore their cosmological and quantum properties.

To elucidate how Milne-like spacetimes extend through the big bang, we give the following brief overview on the subject.

\subsection{The Big Bang Singularity}

The first step in developing a cosmological theory  is to assume the Copernican principle. This assumption is supported by the highly uniform CMB radiation. The Copernican principle implies that the spacetimes, $(M,g)$, which model cosmology are 
\begin{equation}
M = I \times \S \:\:\:\: \text{ and } \:\:\:\: g = -d\tau^2 + a^2(\tau) h 
\end{equation}
where $I\subset \R$ is an interval and $(\S,h)$ are spaces of constant sectional curvature (i.e. maximally symmetric spaces). These are called \emph{FLRW spacetimes}. Let $\tau_{\text{initial}} = \inf I$. If one assumes the universe is in a radiation-dominated era for all $\tau_{\text{initial}} < \tau < \tau_0$ given some $\tau_0$, then one finds $a(\tau) \to 0$ as $\tau \searrow \tau_{\text{initial}}$ and $\tau_{\text{initial}} > -\infty$. In this case we say $\tau_{\text{initial}}$ is the \emph{big bang}. By shifting coordinates, we can assume $\tau_{\text{initial}} = 0$. Moreover the scalar curvature diverges as $\tau \searrow 0$, so $\tau_{\text{initial}} = 0$ admits a \emph{curvature singularity}. Therefore the big bang is labeled as a genuine singularity.

Moreover these arguments generalize if one replaces the assumption that the universe is in a radiation-dominated era with the assumption that the universe obeys the \emph{strong energy condition}.  However there is an exceptional case. This is provided by the classical \emph{Milne universe} where $a(\tau) = \tau$ and $\S = \R^3$ and $h$ is the hyperbolic metric with constant sectional curvature $-1$. In this case the big bang, $\tau = 0$, is just a \emph{coordinate singularity}. But the classical Milne universe is isometric to a proper subset of Minkowski space, so it corresponds to an expanding universe with no energy/matter in it. This is not physically relevant because our universe clearly has matter in it.

The singularity theorems of Hawking and Penrose \cite{HE} demonstrated  that singularities (in the sense of timelike or null geodesic incompleteness) are a generic feature of physically relevant spacetimes. These theorems don't assume any symmetry conditions on the spacetime manifold, but they do assume the strong energy condition. Hawking's cosmological singularity theorem \cite[Theorem 55A]{ON} applies to the Milne universe. However, since the Milne universe embeds into Minkowski space, the past incomplete geodesics are merely a consequence of the past Cauchy horizon given by the future lightcone at the origin of Minkowski space. Hawking's other cosmological singularity theorem \cite[Theorem 55B]{ON} holds only for compact spacelike slices. However it  generalizes to the noncompact case via \cite[Theorem 3.1]{galloway_1986}. The Milne universe does not apply because no spacelike slice in the Milne universe is past causally complete with respect to the origin in Minkowski space.

There is a problem with the strong energy condition assumption in the singularity theorems. Assuming this condition in our universe, one finds that the particle horizon is finite. This implies that there are parts of the CMB that never achieved causal contact in the past. But if this is true, then how could the CMB have such a perfectly uniform temperature? This became known as the \emph{horizon problem} \cite{WeinbergCos}. However there is an exceptional scenario. The Milne universe satisfies the strong energy condition, but there is no horizon problem. The particle horizon is infinite: $\int_0^1\frac{1}{a}d\tau = \int_0^1\frac{1}{\tau}d\tau = +\infty$. But this counterexample is not physically relevant because our universe is not modeled by the Milne universe.

A resolution to the horizon problem is to assume that the universe underwent a brief period of accelerated expansion, $a''(\tau) > 0$, immediately after the big bang and right before the radiation-dominated era. This would allow for causal contact between the different points on the CMB. This theory became known as \emph{inflationary theory} and was first put forth by Alan Guth \cite{GuthInflation}. It also solved the \emph{flatness problem} of cosmology and the \emph{magnetic monopole problem} of certain grand unified theories \cite{WeinbergCos}.

Assuming an inflationary era, $a''(\tau) > 0$, then Friedmann's equations imply that the strong energy conditon \emph{must be violated}. Therefore the singularity theorems above no longer apply. New singularity theorems were sought that did not require the strong energy condition. This was done by Borde and Vilenkin \cite{Borde, BV} and others.\footnote{See \cite{GLcosmo} for a connection between singularities and topology without the strong energy condition.} It was found that some models of inflationary theory also violate the weak energy condition \cite{BVweakviolation}. Then Guth, Borde, and Vilenkin produced a singularity theorem \cite{BGV}, which showed that, even if the weak energy condition is violated, then one has past incompleteness. However their theorem only applies to \emph{inflating regions} of a spacetime. An example of a spacetime with an inflating region is Minkowski space with the inflating region being the Milne universe. Their theorem applies to the Milne universe (because it only requires an averaged Hubble expansion condition), but the conclusion is simply that the inflating region ends at the origin's future lightcone, i.e. at the past Cauchy horizon of the Milne universe.

Since the Milne universe seems to keep offering various counterexamples, it is worth exploring if there are spacetimes \emph{like} the Milne universe which also extend through the big bang. Consider the following scenario: Let $\e > 0$ be a really small number and suppose $a(\tau) = \tau$ for all $0 < \tau < \e$. Then this is modeled by a small portion of the Milne universe (which we know extends through the big bang). Now for $\e \leq \tau < \infty$, imagine $a(\tau)$ smoothly transitions to a radiation-dominated era, then smoothly transitions to a matter-dominated era, and then smoothly transitions to a dark energy-dominated era. Then this spacetime would model the dynamics of our universe and yet \emph{it extends} through the big bang since it was just the Milne universe for $0 < \tau < \e$. Moreover this spacetime solves the horizon problem since $\int_0^\e\frac{1}{a} d\tau= \int_0^\e\frac{1}{\tau}d\tau = +\infty$. 

The physically interesting quality of Milne-like spacetimes is that they generalize what happened in this scenario. Extensions through the big bang, $\tau = 0$, exist provided only a \emph{limiting condition} on the scale factor is satisfied. This condition is simply $a(\tau) = \tau + o(\tau^{1 + \e})$ for some $\e > 0$. Moreover one finds that these spacetimes again solve the horizon problem. Other interesting cosmological properties are proved in section \ref{cosmo properties}.

Going back to our scenario above, it's physically unreasonable to have the universe start off as $a(\tau) = \tau$ for $0 < \tau < \e$, because this universe is void of matter/energy, so how could it smoothly transition to a radiation-dominated era? Instead use the inflationary spacetime, $a(\tau) = \sinh(\tau)$. This is Milne-like, and so it admits an extension through the big bang, $\tau = 0$. Also $a(\tau) = \sinh(\tau)$ has a positive energy density, $\rho(\tau)$, so this universe has energy in it. Therefore let's assume $a(\tau) = \sinh(\tau)$ for $0 < \tau < \e$, and then smoothly transitions to a radiation-dominated era, and then smoothly transitions to a matter-dominated era, and then smoothly transitions to a dark energy-dominated era. Thus this spacetime models our observable universe with a positive energy density initial condition, and it extends through the big bang, $\tau = 0$.

We also mention that Milne-like spacetimes do not disagree with models of inflationary theory that involve an inflaton scalar field $\phi$. Milne-like spacetimes just do not require a scalar field $\phi$ in their definition. It is entirely possible to incorporate an inflaton scalar field in a Milne-like spacetime. In fact, it's shown in section \ref{cosmo properties} that there is a correlation between the initial conditions of the energy density/pressure function of a Milne-like spacetime and inflaton scalar fields $\phi$ in a slow-roll potential $V(\phi)$. 

\subsection{Summary of Results}

Let $(\R^3,h)$ be hyperbolic space with curvature $-1$. Then $(M,g)$ is \emph{Milne-like} if $M = (0,\tau_{\text{max}}) \times \R^3$ and $g = -d\tau^2 + a^2(\tau)h$ where $a(\tau) = \tau + o(\tau^{1 + \e})$ for some $\e > 0$. The metric for Milne-like spacetimes can be written as
\begin{align}
g &= -d\tau^2 + a^2(\tau)\big[dR^2 + \sinh^2(R)(d\theta^2 + \sin^2\theta d\phi^2)]
\\
&= \Omega^2(\tau)[-dt^2 + dx^2 + dy^2 + dz^2]
\end{align}
with respect to two different sets of coordinates $(\tau, R, \theta, \phi)$ and $(t,x,y,z)$. For the classical Milne universe (i.e. $a(\tau) = \tau$), we have $\Omega = 1$.  The assumption $a(\tau) = \tau + o(\tau^{1 + \e})$ implies $\Omega(0)$ is a positive number. This is what allows us to extend these spacetimes through the big bang, $\tau = 0$. Therefore the big bang is just a \emph{coordinate singularity} for Milne-like spacetimes. This is analogous to how the $r = 2m$ event horizon in the Schwarzschild metric is a coordinate singularity. 

We determine the symmetry group for Milne-like spacetimes to be $\R \times \text{O}(1,3)$. We propose that this should be the symmetry group for quantum theory.  Before explaining how we arrive at this symmetry group, let's contrast this with the Poincar{\'e} group, $\R^4 \rtimes \text{O}(1,3)$. What are the symmetry properties of the Poincar{\'e} group? They are the isometries on Minkowski space. Since symmetries are represented by automorphisms on a projective separable complex Hilbert space $\text{P}\mc{H}$, one makes the following assumption when developing quantum theory on Minkowski space.

\medskip

\noindent{\bf Relativistic Invariance Postulate.} \emph{There is a projective unitary representation of the Poincar{\'e} group into the automorphism group of $\emph{\text{P}}\mc{H}$. }

\medskip

Then Bargmann's Theorem \ref{Bargmann Theorem} implies that any projective unitary representation coming from the simply connected double cover, $\R^4 \rtimes \text{SL}(2,\C)$, of the identity component of $\R^4 \rtimes \text{O}(1,3)$ can be lifted to a unitary representation. The irreducible unitary representations are interpreted as elementary particles. The eigenvalues of the two Casimir operators specify the particle's mass and spin. This classification was first done by Wigner \cite{wigner_classification}. 

Although the classification of the Poincar{\'e} group is successful in describing the mass and spin properties of elementary particles, it also describes particles which travel faster than light, particles with negative energy, and a family of zero mass particles which have never been observed \cite{Weinberg, FollandQFT}. Moreover there is a fundamental problem with using the Poincar{\'e} group as the symmetry group for quantum theory.
\[
\text{\emph{Our universe is not modeled by Minkowski space}}.
\]
If our universe is not modeled by Minkowski space, then why use its isometry group as the symmetry group for quantum theory? Of course Minkowski space locally approximates our small neighborhood of the universe, but how does this approximation fit in with the irreducible unitary representations of the Poincar{\'e} group? Does the electron's mass and spin change if the Minkowski approximation fails?

\begin{figure}[h]
\[
\begin{tikzpicture}[scale = 0.5]

\shadedraw [dashed, thick, blue] (-4,2) -- (0,-2) -- (4,2);

\shadedraw[top  color = white, bottom color = gray, dashed, thick, blue] (-4,-6) -- (0,-2) -- (4,-6);
	
\draw [<->,thick] (0,-6.5) -- (0,2.5);
\draw [<->,thick] (-4.5,-2) -- (4.5,-2);

t\draw (.3550,2.90) node [scale = .85] {$t$};
\draw (4.90, -2.25) node [scale = .85] {$x^i$};

\draw [->] [thick] (-2.25,3) arc [start angle=60, end angle=0, radius=50pt];

\draw (-3.1,3.25) node [scale = .85] {\small{$I^+(\ms{O})$}};

\draw [thick, red] (-3.5,2) .. controls (0, -1.3).. (3.5,2);
	
\draw [->] [thick] (2.25,-6.5) arc [start angle=-150, end angle=-180, radius=80pt];	

\draw (2.75,-7.00) node [scale = .85] {\small{$I^-(\ms{O})$}};

\draw [->] [thick]  (1.3,-1.5) arc [start angle=-90, end angle=-180, radius=25pt];
\draw (3.5,-1.5) node [scale = .85] {\small{$\tau =$ constant }};

\end{tikzpicture}
\]
\caption{\small{ A PT extension. The origin is denoted by $\ms{O}$. The original Milne-like spacetime is $I^+(\ms{O})$. It represents our universe. The slices of constant $\tau$ are hyperboloids which foliate $I^+(\ms{O})$. The `spacetime mirror' is $I^-(\ms{O})$. It's isometric to $I^+(\ms{O})$ under the isometry $(t,x,y,z) \mapsto (-t,-x,-y,-z)$.  }}\label{matter-antimatter intro}
\end{figure}
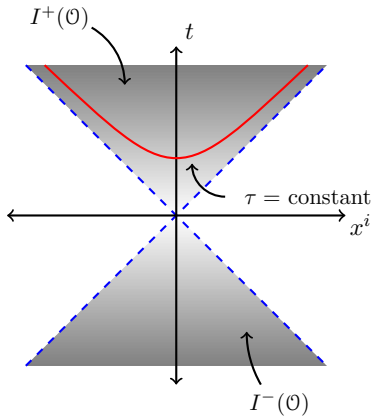

We believe that the symmetry group for quantum theory should come from the symmetries of the observable universe. 
Since Milne-like spacetimes \emph{can} model our observable universe, we propose that the symmetry group come from the symmetric properties of Milne-like spacetimes. First, the group of isometries which fix the origin $\ms{O}$ coincides with the Lorentz group, $\text{O}(1,3)$. This is Theorem \ref{milne isometry}. We have dubbed these isometries \emph{$\ms{O}$-fixing causal isometries}. The factor $\R$ in $\R \times \text{O}(1,3)$ represents \emph{cosmic time translations.} These are maps which shift constant $\tau$ slices, e.g. $(\tau, R, \theta, \phi) \mapsto (\tau + \tau_0,R, \theta, \phi)$. Physically, they shift each comoving observer along its timelike geodeisc by an amount $\tau_0$. Cosmic time translations commute with the $\ms{O}$-fixing causal isometries. The reason why we have $\R$ and not $[0,\infty)$ is because we exclusively work with the PT extension of Milne-like spacetimes. These extensions produce a `spacetime mirror' of our universe. The original Milne-like spacetime is $I^+(\ms{O})$ while the spacetime mirror is $I^-(\ms{O})$. This is illustrated in Figure \ref{matter-antimatter intro}.

 Cosmic time translations commute with $\ms{O}$-fixing causal isometries. Hence the group formed by taking compositions of $\ms{O}$-fixing causal isometries and cosmic time translations is the direct product $\R \times \text{O}(1,3)$. This is our symmetry group. We emphasize that this is a \emph{symmetry in cosmic time}. To explain this point, consider a student learning special relativity for her first time. A natural question for her to ask is: 
 
\medskip 

 \emph{If time is relative for each observer, then how does it make sense to say the universe is 13.8 billion years old?}
 
\medskip 
 
The answer is that it's just a property of FLRW spacetimes, i.e. it's a consequence of the Copernican principle. FLRW spacetimes yield a preferred set of observers called \emph{comoving observers.} When we say the universe is 13.8 billion years old, what we really mean is that each comoving observer has experienced a proper time of 13.8 billion years. There is an inherent symmetry within the universe when one says that all the comoving observers in a FLRW spacetime experience the same proper time. This symmetry should correspond to some cosmic invariance. Theorem \ref{cosmic time invariance} shows that elements from $\R \times \text{O}(1,3)$ are \emph{isometries in the cosmic time direction}. We believe this is the cosmic invariance just described. This is our justification for saying  $\R \times \text{O}(1,3)$ is the symmetry group for Milne-like spacetimes.  Then analogous to the relativistic invariance postulate, we formulate the

\medskip
\noindent{\bf Cosmological Invariance Postulate.} \emph{There is a projective unitary representation of the Milne-like symmetry group into the automorphism group of $\emph{\text{P}}\mc{H}$. }
\medskip

Bargmann's theorem applies, and so any projective unitary representation coming from the simply connected double cover, $\R \times \text{SL}(2,\C)$, of the identity component of $\R \times \text{O}(1,3)$ can be lifted to a unitary representation.
We classify the irreducible unitary representations. We find

\begin{center}
 \begin{tabular}{c| c |c} 
 
 Orbit $O_m$ & Representative mass $m$ & Little Group $H_m$   \\ [.75ex] 
 \hline
 & & 
 \\
 $O_m^+$ & $|m|$ & $\text{SL}(2,\C)$ 
  \\ [.75ex] 

 $O_m^-$ & $-|m|$ & $\text{SL}(2,\C)$ 
  \\[.75ex] 

 $O_0$ & $0$ & $\text{SL}(2,\C)$ 
 \\ [.75ex] 

\end{tabular}
\end{center}

The mass $m$ corresponds to the eigenvalue of a mass Casimir operator. The little group is $\text{SL}(2,\C)$ for each of the orbits because $\R \times \text{SL}(2,\C)$ is a direct product. This differs from the Poincar{\'e} group because the Poincar{\'e} group is a semi-direct product. Since irreducible unitary representations are associated with elementary particles, this table describes the mass of these particles. An interpretation is: The orbit $O_m^+$ corresponds to particles with positive mass with respect to the future-pointing $\pd_\tau$. These particles make up $I^+(\ms{O})$. The orbit $O_m^-$ corresponds to particles with negative mass with respect to $\pd_\tau$. Hence it corresponds to particles with positive mass with respect to past-pointing $-\pd_\tau$. These particles make up $I^-(\ms{O})$. Therefore ``negative energy" can be interpreted as ``positive energy traveling in the $-\pd_\tau$ direction." The orbit $O_0$ corresponds to massless particles which move at the speed of light. Each of the three distinctive orbits $O^+_m$, $O^-_m$, and $O_0$ has a physical interpretation. This is unlike the classifaction for the Poincar{\'e} group. The majority of the orbits for the Poincar{\'e} group lack any real physical meaning.

To finish the classification, one has to know the irreducible unitary representations of $\text{SL}(2,\C)$. This was done independently by Bargmann \cite{Barg2} and Gelfand and Naimark \cite{GelfandNaimark}.  Bargmann found that each irreducible representation has an associated spin $j = 0,\: 1/2,\: 1, \dotsc$. However there are 3 Casimir operators for $\R \times \text{SL}(2,\C)$ as oppose to just two for $\R^4 \rtimes \text{SL}(2,\C)$. It would be interesting if there are any physical consequences of this. For example spin 0 particles (e.g. the Higgs boson) are in a different class than half-integer spin particles (e.g. the electron).  

Since the mass Casimir operator is generated from cosmic time translations, we postulate that the mass operator corresonds to $i\pd_\tau$ (in $\hbar = 1$ units) and that spinor fields $\psi$ on the PT extension of a Milne-like spacetime satisfy

\begin{equation}\label{fundamental eq intro}
i\pd_\tau \psi = m\psi.
\end{equation}
The \emph{Dirac equation}
 for Dirac spinor fields $\psi$ on Milne-like spacetimes is
\begin{equation}\label{dirac intro}
\big[\Omega(\tau)\g^\mu\pd_\mu\big] \psi = m\psi.
\end{equation}
For the classical Milne universe (i.e. $\Omega = 1$),  we reproduce the original Dirac equation on Minkowski space. Choosing coordinates $(t,x,y,z)$ which align with a comoving observer shows that equations (\ref{fundamental eq intro}) and (\ref{dirac intro}) agree.  Thus

\[
\emph{\text{
The Dirac equation naturally corresponds to cosmic time translations.
}}
\]

\noindent This gives more credence that cosmic time translations should be part of the symmetry group for quantum theory.

 Since the Dirac equation on Milne-like spacetimes is very similar to the original Dirac equation on Minkowski space, the usual properties hold. There is still a conserved probably current. Lorentz invariance holds because the original Dirac equation is Lorentz invariant and $\Omega(\tau) = \Omega(\Lambda \tau)$ for any $\ms{O}$-fixing causal isometry $\Lambda \in \text{O}(1,3)$. 
 
 When solving the Dirac equation, we distinguish between $I^+(\ms{O})$ and its PT isometric image $I^-(\ms{O})$. We say $\psi$ \emph{solves the Dirac equation for} $I^+(\ms{O})$ if 
\begin{equation} 
 \big[\Omega \g^\mu \pd_\mu\big] \psi = m\psi.
 \end{equation}
 We say $\psi$ \emph{solves the Dirac equation for} $I^-(\ms{O})$ if 
\begin{equation} 
 \big[\Omega \g^\mu (-\pd_\mu)\big] \psi = m\psi.
 \end{equation}
 The idea is that observers would use coordinates $(t,x,y,z)$ in  $I^+(\ms{O})$  while observers in $I^-(\ms{O})$ would use $(-t,-x,-y,-z)$ as their coordinates. Note that solving for $I^-(\ms{O})$ is equivalent to solving $\big[\Omega(\tau)\g^\mu\pd_\mu\big] \psi = -m\psi.$
 
 We introduce electromagnetism into the Dirac equation via an electromagnetic potential $A_\mu$ and investigate how the Dirac equation transforms under $\text{PT}$-reversal and complex conjugation. The results are

\begin{center}
 \begin{tabular}{c| c |c} 
 
 Spinor field & Equation & An interpretation   \\ [.75ex] 
 \hline
 & & 
 \\
 $\psi$ & $\Omega \g^\mu(\pd_\mu + ieA_\mu)\psi = m\psi$ & $\psi$ in $I^+\ms{O})$
  \\ [.75ex] 

 $\psi^*$ & $\Omega \g^\mu(-\pd_\mu + ieA_\mu)\psi^* = m\psi^*$ & $\psi$ in $I^-(\ms{O})$ 
  \\[.75ex] 

 $\text{PT}\psi$ & $\Omega \g^\mu(-\pd_\mu - ieA_\mu)\text{PT}\psi = m\text{PT}\psi$ & Anti $\psi$ in $I^-(\ms{O})$  
 \\ [.75ex] 
 
 $\text{PT}\psi^*$ & \:\: $\Omega \g^\mu(\pd_\mu - ieA_\mu)\text{PT}\psi^* = m\text{PT}\psi^*$ \:\: & Anti $\psi$ in $I^+(\ms{O})$ 
 \\ [.75ex] 

\end{tabular}
\end{center}

Given the interpretation, we believe the relationship between $\psi$ and $\text{PT}\psi$  suggests that our universe's missing antimatter comprises $I^-(\ms{O})$.

\medskip

\noindent{\bf In Conclusion:}

\begin{itemize}

\item[(1)] We show that Milne-like spacetimes extend beyond the big bang. The extension can contain a universe, $I^-(\ms{O})$, which is isometric to our universe, $I^+(\ms{O})$, under time and parity reversal.

\item[(2)] We argue that the symmetry group for quantum theory should come from the Milne-like symmetry group, $\R \times \text{O}(1,3)$, and not from the Poincar{\'e} group, $\R^4 \rtimes \text{O}(1,3)$. This is because our universe can be modeled by a Milne-like spacetime. Our universe cannot be modeled by Minkowski space. 

\item[(3)] The irreducible unitary representations for the Milne-like symmetry group can be interpreted as elementary particles with mass and spin. Unlike the classification for the Poincar{\'e} group, we do not obtain any faster-than-light particles. Moreover each distinctive orbit $O^+_m$, $O^-_m$, and $O_0$ has a physical interpretation. This is not true for the Poincar{\'e} group. The majority of the orbits in the Poincar{\'e} group lack any real physical meaning. Perhaps there is a physical reason to choose $\R^4 \rtimes \text{O}(1,3)$ as oppose to $\R \times \text{O}(1,3)$ as the symmetry group, but we have not found one.

\item[(4)] The mass Casimir operator is generated from cosmic time translations. This yields an eigenvalue problem 
\begin{equation}
i\pd_\tau \psi = m \psi
\end{equation}
 for spinor fields $\psi$ on the PT extension of a Milne-like spacetime. For Dirac spinors this equation is the same as the Dirac equation. Therefore the Dirac equation naturally corresponds to cosmic time translations. This gives more credence that cosmic time translations should be used in the symmetry group for quantum theory.

\item[(5)] PT transformations on $\psi$ in $I^+(\ms{O})$ yield an anti $\psi$ in $I^-(\ms{O})$.  We believe that this fact along with the CPT theorem \cite{PCT} supports the claim that the universe's missing antimatter comprises $I^-(\ms{O})$.

\end{itemize}

\newpage

\section{Milne-like Spacetimes and their Symmetries}\label{milne section}

\subsection{Cosmological Properties}\label{cosmo properties}

A \emph{spacetime} is a differentiable manifold $M$ equipped with a nondegenerate continuous Lorentzian metric $g$ such that $(M,g)$ is time-oriented. $(M,g)$ \emph{extends} if there is a spacetime $(M_\ext, g_\ext)$, of the same dimension, such that $(M,g)$ embeds isometrically as a proper subset of $(M_\ext, g_\ext)$.

\medskip

\begin{Def}\label{milne def}
\emph{
A spacetime $(M,g)$ is  \emph{Milne-like} if
\begin{equation}
M = (0,\tau_{\max}) \times \R^3\:\:\:\: \text{ and } \:\:\:\: g = -d\tau^2 + a^2(\tau)h
\end{equation}
 where $a(\tau) = \tau + o(\tau^{1+\e})$ for some $\e > 0$, and $(\R^3,h)$ is the complete hyperbolic space with constant sectional curvature $-1$. We call $\tau$ the  \emph{cosmic time} and $a(\tau)$ the \emph{scale factor}. The curves $\tau \mapsto (\tau, R_0, \theta_0, \phi_0)$ are the \emph{comoving observers}. 
}
\end{Def}

\noindent\emph{Remarks.} 

\begin{itemize}

\item[$\bullet$] The condition $a(\tau) = \tau  + o(\tau^{1+\e})$ means that $a(\tau) = \tau + f(\tau)$ where $f(\tau)/\tau^{1+\e} \to 0$ as $\t \searrow 0$.  Since the definition only requires a limiting condition on the scale factor, Milne-like spacetimes are very robust. One can construct a scale factor which starts off inflationary, then smoothly transitions to a radiation-dominated era,  then smoothly transitions to a matter-dominated era,  and then smoothly transitions to a dark energy-dominated era. Then $a(\tau)$ would correspond to a Milne-like spacetime modeling the dynamics of our universe. 

\item[$\bullet$] If the sectional curvature is left arbitrary, then one only needs to rescale $a(\tau)$ to obtain a Milne-like spacetime. If we restore factors of $c$, then the sectional curvature has to be set to $-c^{-2}$. In this case a spacetime is Milne-like if $a(\tau) = \frac{1}{c}\tau + o(\tau)^{1 + \e}$. If our universe really is modeled by a Milne-like spacetime, then this establishes a connection between the spatial curvature of our universe and the speed of light.

\end{itemize}

\medskip

The proof of the following theorem shows that the big bang, $\tau = 0$, is just a coordinate singularity for Milne-like spacetimes. This is analogous to how the $r = 2m$ event horizon in the Schwarzschild metric is just a coordinate singularity.

\medskip

\begin{thm}[\cite{GalLing_con}]\label{extension thm}
Milne-like spacetimes extend.
\end{thm}

\proof
 In coordinates $(\tau, R, \theta, \phi)$ the metric can be written as 
\begin{equation}
g = -d\tau^2 + a^2(\tau)\big[dR^2 + \sinh^2(R)(d\theta^2 + \sin^2\theta d\phi^2) \big]
\end{equation}
Fix $0 < \tau_0 < \tau_{\text{max}}$. Define new coordinates $(t,r,\theta, \phi)$ by

\begin{equation}\label{t and r}
t =b(\tau)\cosh(R ) \:\:\:\: \text{ and } \:\:\:\: r = b(\tau)\sinh(R)
\end{equation}
where
\begin{equation}
b(\tau) = \exp\left(\int_{\tau_0}^\tau \frac{1}{a(s)}ds \right).
\end{equation}
Therefore 
\begin{equation}\label{def for tau}
\tau = b^{-1}\big( \sqrt{t^2 - r^2}\big).
\end{equation}
With respect to these coordinates, the metric takes the form 
\begin{align}
g &= \Omega^2\big(\tau(t,r)\big)\big[-dt^2 + dr^2 + r^2(d\theta^2 + \sin^2\theta d\phi^2)\big]. 
\end{align}
where
\begin{equation}
\Omega(\tau) = \frac{1}{b'(\tau)} = \frac{a(\tau)}{b(\tau)}.
\end{equation}

It suffices to show $\Omega(0) := \lim_{\tau \searrow 0}\Omega(\tau)$ exists and $0 < \Omega(0) < \infty$. Because if this is true, then there is no degeneracy in the metric at $\tau = 0$ in these coordinates. Therefore one can extend the metric through $\tau = 0$.

To show $0 < \Omega(0) < \infty$, put $b'(0) = \lim_{\tau \searrow 0}b'(\tau) = \lim_{t\searrow 0} b(\tau)/a(\tau)$. Let $f(\tau) = a(\tau) - \tau$. By assumption there is an $\a > 0$ such that  $\lim_{\tau \searrow 0}f(\tau)/\tau^{1 + \a} = 0$. Therefore for any $\e > 0$, there exists a $\delta > 0$ such that for all $0 < \tau < \delta$, we have $|f(\tau)| < \e\tau^{1 + \a}$. Hence $\tau - \e\tau^{1 + \a} < \tau + f(\tau) < \tau + \e\tau^{1 + \a}$. Therefore $b(\tau)/a(\tau)$ is squeezed between 
\begin{equation}
\frac{1}{a(\tau)}\exp\left(-\int_{\tau}^{\tau_0}\frac{1}{(\tau - \e\tau^{1 + \a})}ds \right) < \frac{b(\tau)}{a(\tau)} < \frac{1}{a(\tau)}\exp\left(-\int_\tau^{\tau_0}\frac{1}{(\tau + \e\tau^{1 + \a})}ds \right)
\end{equation}
Evaluating the integrals we find 
\begin{equation}
\frac{1}{\tau_0}\left(\frac{\tau}{a(\tau)} \right)\left( \frac{1 - \e\tau^\a}{1 + \e\tau_0^\a}\right)^{-1/\a} < \frac{b(\tau)}{a(\tau)} < \frac{1}{\tau_0}\left(\frac{\tau}{a(\tau)} \right)\left( \frac{1 + \e\tau^\a}{1 + \e\tau_0^\a}\right)^{-1/\a}
\end{equation}
Since this holds for all $0 < \tau < \delta$, we have $\Omega(0) = 1/b'(0) = \tau_0$.

\qed

\medskip 

\noindent\emph{Remark.} Throughout this paper, we will use $(\tau, R, \theta, \phi)$ for coordinates on a Milne-like spacetime and $(t,r, \theta, \phi)$ or $(t,x,y,z)$ for the extension.

 \medskip

\begin{figure}[h]
\[
\begin{tikzpicture}[scale = 0.75]

\shadedraw [dashed, thick, blue](-4,2) -- (0,-2) -- (4,2);

\draw [<->,thick] (0,-3.5) -- (0,2.25);
\draw [<->,thick] (-4.5,-2) -- (4.5,-2);

\draw (.35,2.5) node [scale = .85] {$t$};
\draw (4.75, -2.25) node [scale = .85] {$x^i$};
\draw (-.25,-2.25) node [scale = .85] {$\ms{O}$};


\draw [->] [thick] (1.5,2.8) arc [start angle=120, end angle=180, radius=40pt];
\draw (2.0,3.05) node [scale = .85]{\small{$I^+(\ms{O})$}};

\draw [->] [thick] (-2.2,-1.8) arc [start angle=-90, end angle=-30, radius=40pt];
\draw (-3.1,-1.7) node [scale = .85] {$\pd J^+(\ms{O})$};


\draw [thick, red] (-3.84,2) .. controls (0,-2) .. (3.84,2);
\draw [thick, red] (-3.5,2) .. controls (0, -1.3).. (3.5,2);

\draw [->] [thick]  (1,-2.3) arc [start angle=-120, end angle=-180, radius=40pt];
\draw (2.3,-2.5) node [scale = .85] {\small{$\tau =$ constant }};

\end{tikzpicture}
\]
\caption{\small{A Milne-like spacetime embedded in a larger spacetime. The Milne-like spacetime is given by $I^+(\ms{O})$. The constant $\tau$-slices are hyperboloids which foliate $I^+(\ms{O})$.}}\label{milne-like figure}
\end{figure}
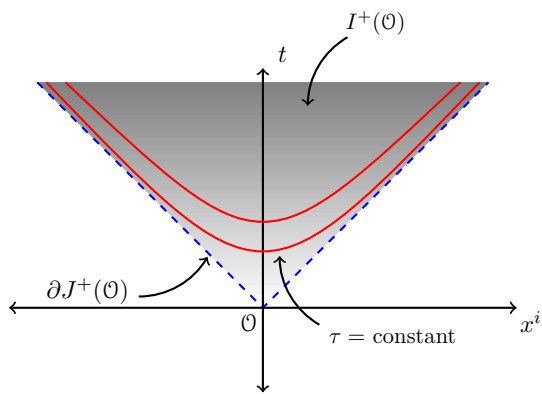

 Recall that for a spacetime $(M,g)$ and point $p \in M$, the \emph{causal future} $J^+(p)$ is the set of points $q \in M$ such that there is a future directed causal curve connecting $p$ to $q$. The \emph{causal past} $J^-(p)$ is defined by switching future to past. The \emph{timelike future} and \emph{timelike past} $I^\pm(p)$ are defined by switching causal to timelike. Physically, $J^+(p)$ represents the subset of spacetime which can be influenced by $p$. We let $\ms{O}$ denote the origin of our extension, (i.e. $t(\ms{O}) = r(\ms{O}) = 0$). Then the Milne-like spacetime coincides with $I^+(\ms{O})$ while the lightcone coincides with $\pd J^+(\ms{O})$. See Figure \ref{milne-like figure}.
 
\medskip

\medskip

We provide a list of cosmological properties we were able to deduce about Milne-like spacetimes. These properties show that Milne-like spacetimes are physically interesting.  For example the fifth property may have something to do with dark energy.


\newpage

\noindent {\bf Cosmological Properties}

\begin{itemize}

\item [(i)] Milne-like spacetimes admit continuous extensions through the big bang.

\item[(ii)] There exist examples of inflationary Milne-like spacetimes which admit \emph{smooth} extensions through the big bang, i.e. $g_\ext$ can be chosen to be smooth. Hence there are \emph{no} curvature singularities at the big bang for these spacetimes.

\item[(iii)] The comoving observers of a Milne-like spacetime all emanate from the origin $\ms{O}$ in the extension.

\item[(iv)] Milne-like spacetimes solve the horizon problem of inflationary theory.

\item[(v)] Let $\rho(\tau)$ and $p(\tau)$ be the energy density and pressure function for a smooth Milne-like spacetime. If $a''(0) = 0$ and $a'''(0)$ is finite, then $\rho(0) = - p(0)$. Hence $\rho(0)$ and $p(0)$ have the same form as a cosmological constant.

\item[(vi)] If the energy-momentum tensor is dominated by an inflaton scalar field $\phi$ in a potential $V(\phi)$, then the assumptions ``$a''(0) = 0$ and $a'''(0)$ is finite" from (v) naturally induce an era of slow-roll inflation.

\end{itemize}

\proof

\text{ }

\begin{itemize}

\item[(i)] This is Theorem \ref{extension thm}.

\item[(ii)] An example is $a(\tau) = \sinh(\tau)$. Then $b(\tau) = \tanh(\tau/2)$ and $\Omega\big(\tau(t,r)\big) = 2/(1 - t^2 + r^2)$. Therefore $\Omega \in C^\infty$. Since the metric is infinitely differentiable at the big bang, any scalar produced from the curvature tensor must have a finite valued quantity at $\tau = 0$.

\item[(iii)] A comoving observer, $\tau \mapsto (\tau, R_0, \theta_0, \phi_0)$, is parameterized by $t(\tau) = b(\tau)\cosh(R_0)$ and $r(\tau) = b(\tau)\sinh(R_0)$. Therefore $t(\tau) =C r(\tau)$ for some $C > 1$.

\item[(iv)] Since $a'(0) := \lim_{\tau \searrow 0}a(\tau)/\tau = 1$, for any $\e > 0$ there exists a $\delta > 0$ such that $|a(\tau)/\tau - 1| < \e$ for all $0 < \tau < \delta$. Hence $1/a(\tau) > 1/(1+\e)\tau$ for all $0 < \tau < \delta$, and so $\int_0^\delta d\tau/a(\tau) = + \infty$. Thus the particle horizon is infinite. Alternatively, Figure \ref{milne-like figure} along with point (iii) shows that $J^-(p) \cap J^-(q) \neq \emptyset$ for all points $p$ and $q$ in $I^+(\ms{O})$. Hence all points were causally connected at some point in the past.

\item[(v)] This is an $\e$-$\delta$ argument applied to the Friedmann equations. Let $T = \text{Ric} - \frac{1}{2}R_{\text{sca}}g$ be the energy-momentum tensor. Then the energy density is $\rho = T(\pd_\tau, \pd_\tau)$ and the pressure function is $p = T(e,e)$ where $e$ is any unit spacelike vector orthogonal to $\pd_\tau$. The isotropy of the spatial slices implies $p$ is independent of the chosen $e$. Put $\rho(0) = \lim_{\tau \searrow 0}\rho(\tau)$ and likewise for $p(0)$. Let $f(\tau) = a(\tau) - \tau$. Friedmann's equations are (in $G = 1$ units)
\[
\frac{8\pi}{3} \rho(\tau) = \left(\frac{a'(\tau)}{a(\tau)}\right)^2 - \frac{1}{a(\tau)^2} = \frac{2f'(\tau) + f'(\tau)^2}{\big[\tau + f(\tau)\big]^2} = \frac{\big(f'(\tau)/\tau\big)\big[2/\tau + f'(\tau)\big]}{\big(1 + f(\tau)/\tau\big)^2}
\]
and
\[
-8\pi p(\tau) = 2\frac{a''(\tau)}{a(\tau)} + \frac{8\pi}{3}\rho(\tau) = \frac{2f''(\tau)/\tau}{1 + f(\tau)/\tau} + \frac{8\pi}{3}\rho(\tau).
\]

By definition of a Milne-like spacetime, we have $f'(0) := \lim_{\tau \searrow 0}f(\tau)/\tau = 0$. Since $0 = a''(0) = f''(0) = \lim_{\tau \searrow 0}f'(\tau)/\tau$ and $\a := a'''(0) = \lim_{\tau \searrow 0}f''(\tau) /\tau$, for all $\e > 0$, there is a $\delta > 0$ such that $|f''(\tau)/\tau - \a| < \e$ for all $0 < \tau < \delta$. Integrating this expression gives $(\a - \e)\tau/2 < f'(\tau)/\tau < (\a + \e)\tau/2$. Plugging this into the first Friedmann equation yields $8\pi\rho(0)/ 3 = \a$. Using this for the second Friedmann equation yields $-8\pi p(0) = 3\a$.

\item[(vi)] The energy-momentum tensor for an inflaton scalar field $\phi$ is 
\begin{equation}
T_\phi = d\phi \otimes d\phi - \left[\frac{1}{2}(\nabla \phi, \nabla \phi) + V(\phi) \right]g
\end{equation}
where $V(\phi)$ is the potential of $\phi$. The isotropy of the spatial slices implies $\phi$ is solely a function of $\tau$. Therefore its energy density is given by $\rho_\phi(\tau) = \frac{1}{2}\phi'(\tau)^2 + V\big(\phi(\tau)\big)$. If the assumptions from (v) hold, then we have $\rho_\phi(0) = 3\a/8\pi$ where $\a := a'''(0)$. Therefore for $\tau$ sufficiently small, we have 
\begin{equation}
\frac{1}{2}\phi'(\tau)^2 + V\big(\phi(\tau)\big) \approx \frac{3\a}{8\pi}
\end{equation}
By choosing $V(\phi(\tau)\big) \approx 3\a/8\pi$ for all $\tau$ sufficiently small, we have $\phi'(\tau) \approx 0$. Hence this gives an era of slow-roll inflation.


\end{itemize}

\qed

\medskip

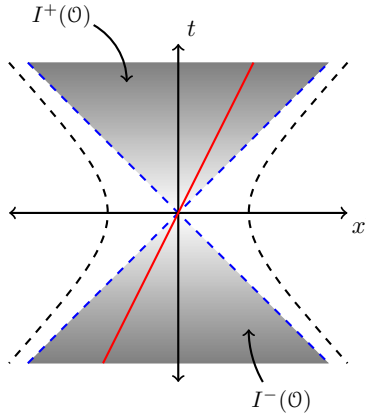
\begin{figure}[h]
\[
\begin{tikzpicture}[scale = 0.5]

\shadedraw [dashed, thick, blue] (-4,2) -- (0,-2) -- (4,2);

\shadedraw[top  color = white, bottom color = gray, dashed, thick, blue] (-4,-6) -- (0,-2) -- (4,-6);
	
\draw [<->,thick] (0,-6.5) -- (0,2.5);
\draw [<->,thick] (-4.5,-2) -- (4.5,-2);

\draw [dashed, thick] (-4.5,2) .. controls (-1, -2) .. (-4.5, -6);

\draw [dashed, thick] (4.5,2) .. controls (1, -2) .. (4.5, -6);

t\draw (.3550,2.90) node [scale = .85] {$t$};
\draw (4.90, -2.25) node [scale = .85] {$x^i$};

\draw [->] [thick] (-2.25,3) arc [start angle=60, end angle=0, radius=50pt];

\draw (-3.1,3.25) node [scale = .85] {\small{$I^+(\ms{O})$}};

\draw [->] [thick] (2.25,-6.5) arc [start angle=-150, end angle=-180, radius=80pt];	

\draw (2.75,-7.00) node [scale = .85] {\small{$I^-(\ms{O})$}};

\draw [thick, red] (-2,-6) -- (2,2);
	
\end{tikzpicture}
\]
\caption{\small{ 
A PT extension $(M,g)$. The spacetime manifold $M$ lies between the black dashed lines. It contains $J^+(\ms{O}) \cup J^-(\ms{O})$. We can always choose $M$ to be simply connected. $I^-(\ms{O})$ is isometric to the Milne-like spacetime $I^+(\ms{O})$ under the PT isometry $x^\mu \mapsto -x^\mu$. A comoving observer and its image under the isometry are shown in red.
}}
\label{matter-antimatter}
\end{figure}

\begin{Def}\label{PT def}
\emph{
Let $(M,g)$ be an extension of a Milne-like spacetime. 
}

\begin{itemize}

\item[$\bullet$] \emph{The origin $\ms{O} \in M$ is the unique point such that $t(\ms{O}) = r(\ms{O}) = 0$. Since all the comoving observers emanate from $\ms{O}$, we will often refer to $\ms{O}$ as the \emph{big bang}.}

\item[$\bullet$] \emph{If $\g(\tau)$ is any comoving observer, there is a natural way to extend $I^+(\ms{O})$ so that the derivative of $\g$ remains continuous through the big bang $\ms{O}$. We say $(M, g)$ is a PT \emph{extension} of $I^+(\ms{O})$ provided the map $(t, x,y,z) \mapsto (-t, -x, -y,-z)$ is an isometry. In this case $I^+(\ms{O})$ and $I^-(\ms{O})$ are isometric. See Figure \ref{matter-antimatter}. Given a point $(\tau, R, \theta, \phi) \in I^+(\ms{O})$, we will denote its image under the isometry as $(-\tau, R, \theta, \phi) \in I^-(\ms{O})$. We will exclusively work with PT extensions in this paper.} 

\item[$\bullet$]  \emph{The vector field $\pd/\pd\tau$ is defined on $I^+(\ms{O})\cup I^-(\ms{O})$ and will \emph{always} assumed to be future-pointing (even in $I^-(\ms{O})$). The comoving observers are the integral curves of $\pd/\pd \tau$.}

\item[$\bullet$] \emph{The proof of Theorem \ref{extension thm} shows that $g = \Omega^2(\tau)\eta$ where $\eta$ is the Minkowski metric and $\Omega = 1/b'$. We call $\Omega$ the  \emph{conformal factor}. Note that $\Omega$ is constant on $\tau$ slices. That is, $\Omega(p) = \Omega(q)$ for all $p,q \in \{\tau = \pm\rm{constant}\}$. Also $\Omega(p) = \Omega(\ms{O})$ for all $p \in \pd J^+(\ms{O}) \cup \pd J^-(\ms{O})$, i.e. for all $p$ on the lightcone of $\ms{O}$.}

\end{itemize}

\end{Def}

 P and T stand for parity and time reversal, respectively. Because of the CPT theorem \cite{PCT}, one can't help but speculate that this has something to do with the matter-antimatter asymmetry problem in our universe. 
 
 A PT extension uniquely determines all causally related points from the big bang $\ms{O}$. That is, if two PT extensions differ, then they differ only on points which are spacelike separated from $\ms{O}$. Because of this we will often talk about \emph{the} PT extension. Since any two PT extensions differ on spacelike separated points of $\ms{O}$, all PT extensions define the same $J^+(\ms{O}) \cup J^-(\ms{O})$. Note that $I^+(\ms{O})$ is our original Milne-like spacetime. Tensor fields on $J^+(\ms{O}) \cup J^-(\ms{O})$ are uniquely determined by their restriction to $I^+(\ms{O}) \cup I^-(\ms{O})$ by continuity. Hence they're uniquely determined by the original Milne-like spacetime $I^+(\ms{O})$.

\subsection{The Milne-like Symmetry Group} 

In this section we seek the symmetries of a Milne-like spacetime. Understanding the symmetries of a spacetime is motivated by quantum theory where symmetries are modeled by automorphisms on a projective Hilbert space. First we determine the isometries on Milne-like spacetimes.

\medskip

\noindent\emph{Remark.} To be mathematically precise, we redefine Milne-like spacetimes so that (i) they admit $C^2$ extensions (i.e. $g_\ext$ is $C^2$) and (ii) the domain of $\tau$ is $(0, +\infty)$, i.e. $\tau_{\text{max}} = +\infty$. For example the inflationary spacetime $a(\tau) = \sinh(\tau)$ satisfies both (i) and (ii). If $a(\tau)$ begins as $\sinh(\tau)$ and smoothly transitions to a radiation-dominated era, and then smoothly transitions to a matter-dominated era, and then smoothly transitions to a dark energy-dominated era, then $a(\tau)$ models the dynamics of our universe and satisfies both (i) and (ii). An example of a scale factor not satisfying (i) is $a(\tau) = \tau + \tau^2$ since the scalar curvature diverges to $+\infty$ as $\t \to 0$.\footnote{An interesting consequence of this is that curvature singularities and coordinate singularities can coincide. In other words curvature singularities are not obstructions to spacetime extensions. } An example of a scale factor not satisfying (ii) is $a(\tau) = \sin(\tau)$.

\medskip

\begin{Def}
\emph{
Let $(M,g)$ be a spacetime and $p \in M$ such that $\big[J^+(p) \cup J^-(p)\big] \setminus \{p\}$ is a manifold with boundary. A  \emph{$p$-fixing causal isometry} is a map
\begin{equation}
f \colon J^+(p) \cup J^-(p) \to J^+(p) \cup J^-(p)
\end{equation}
such that }
\begin{enumerate}
\item $f (p) = p$,

\item $f$ \emph{is a diffeomorphism on} $\big[J^+(p) \cup J^-(p)\big] \setminus \{p\}$ ,
\item  $g_q(X,Y) = g_{f(q)}(df X, df Y)$ \emph{for all} $X,Y \in T_qM$ \emph{and} $q \in J^+(p) \cup J^-(p)  $.
\end{enumerate}
\emph{
The set of $p$-fixing causal isometries on $(M,g)$ forms a group under composition. 
}

\end{Def}

\medskip

\begin{thm}\label{milne isometry} Let $(M,g)$ be the \emph{PT} extension of a Milne-like spacetime. 
The group of $\ms{O}$-fixing causal isometries is isomorphic to the Lorentz group, $\emph{\text{O}}(1,3)$.
\end{thm}

\proof 
We will first demonstrate that the Lorentz group is isomorphic to a subgroup of $\ms{O}$-fixing causal isometries on $(M,g)$. Then we will show that the Lorentz group is the whole group.

 Let $\Lambda^\mu_{\:\:\:\nu}$ be an element of the Lorentz group i.e., it satisfies $\Lambda^\alpha_{\:\:\:\mu}\Lambda^\beta_{\:\:\:\nu}\eta_{\alpha\beta} = \eta_{\mu\nu}$ where $\eta_{\mu\nu} = \rm{diag}[-1,1,1,1]$ is the usual Minkowski metric. It produces a unique map, $x \mapsto \Lambda x$, from $J^+(\ms{O})\cup J^-(\ms{O})$ to itself via $x^\mu \mapsto \Lambda^\mu_{\:\:\:\nu}x^\nu$ where $x^\mu$ are the standard $(t,x,y,z)$ coordinates on $M$. Clearly $\Lambda \ms{O} = \ms{O}$. Since $\Lambda$ takes points on $\{\tau = \rm{constant}\}$ slices to $ \{\tau = \pm \rm{constant}\}$ slices and points from the boundary of the cones $\pd J^+(\ms{O}) \cup \pd J^-(\ms{O})$ to itself, we see that $\Lambda$ is a diffeomorphism on $\big(J^+(\ms{O})\cup J^-(\ms{O})\big) \setminus \{\ms{O}\}$. For $p \in J^+(\ms{O})\cup J^-(\ms{O})$ and $X,Y \in T_pM$, we have $d\Lambda X = \Lambda^\mu_{\:\:\:\nu}X^\nu\pd_\mu \in T_{\Lambda p}M$. Therefore
\begin{align}
g_{\Lambda p}(d\Lambda X, d\Lambda Y) &= \Omega^2(\Lambda p)\eta(d\Lambda X, d\Lambda Y) \notag
\\
&= \Omega^2(p) \eta_{\mu\nu}(\Lambda^\mu_{\:\:\:\a}X^\a)(\Lambda^\nu_{\:\:\: \b}Y^\b) \notag
\\
&= \Omega^2(p) \eta(X,Y) \notag
\\
&=g_p(X,Y).
\end{align}
We used $\Omega(\Lambda p) = \Omega(p)$ in the second line since either both $p$ and $\Lambda p$ lie in the same $\tau$ slice or both $p$ and $\Lambda p$ lie on the future or past cones $\pd J^+(\ms{O}) \cup \pd J^-(\ms{O})$. 

Now we show that the Lorentz group is the whole group. Let $f$ be an $\ms{O}$-fixing causal isometry on $(M,g)$. The differential map $df_{\ms{O}}\colon T_{\ms{O}}M \to T_{\ms{O}}M$ is a linear isometry on the tangent space at $\ms{O}$. Therefore $df_{\ms{O}}$ corresponds to an element of the Lorentz group, say $\Lambda^\mu_{\:\:\:\nu}$. It operates on vectors $X \in T_{\ms{O}}M$ via $df(X) = \Lambda^\mu_{\:\:\:\nu}X^\nu\pd_\mu \in T_{\ms{O}}M$. Now define the $\ms{O}$-fixing causal isometry $\tilde{f}$ by $\tilde{f}(x) = \Lambda^\mu_{\:\:\:\nu}x^\nu$. Consider the set 
\begin{equation}
A = \{p \in J^+(\ms{O}) \cup J^-(\ms{O}) \mid df_p = d\tilde{f}_p\}.
\end{equation}
Note that if $df_p = d\tilde{f}_p$, then $f(p) = \tilde{f}(p)$. Hence it suffices to show $A = J^+(\ms{O}) \cup J^-(\ms{O})$. $A$ is nonempty since $\ms{O} \in A$, and $A$ is closed because $df - d\tilde{f}$ is continuous. So since $J^+(\ms{O}) \cup J^-(\ms{O})$ is connected, it suffices to show $A$ is open in the subspace topology. Let $p \in A$ and let $U \subset M$ be a normal\footnote{This is why we needed to redefine Milne-like spacetimes to admit $C^2$ extensions.} neighborhood about $p$. If $q \in U$, there is a vector $X \in T_pM$ such that $\exp_p(X) = q$. Since isometries map geodesics to geodesics, they satisfy the property $f \circ \exp_p = \exp_{f(p)} \circ df_p$ for all points in $U$. Therefore
\begin{equation}
f(q) = f\big(\exp_{p}(X)\big) = \exp_{f(p)}(df_pX) = \exp_{\tilde{f}(p)}(d\tilde{f}_pX) = \tilde{f}\big(\exp_p(X)\big) = \tilde{f}(q).
\end{equation}
Thus $f(q) = \tilde{f}(q)$ for all $q \in U$; hence $df_q = d\tilde{f}_q$ for all $q \in U$. Therefore $A$ is open.
\qed

\medskip \medskip

\noindent\emph{Remark.} Theorem \ref{milne isometry} shows that Lorentz invariance follows from the isotropy of Milne-like spacetimes. The converse of this statement is interesting. What if the isotropy of our universe is a consequence of Lorentz invariance?

\medskip

Next we define cosmic time translations. These are maps which shift each comoving observer along their timelike geodesic. For example, a cosmic translation would shift the red line in Figure \ref{matter-antimatter} along itself. Recall that given a point $(\tau, p) \in I^+(\ms{O}) = (0, \infty) \times \R^3$, its image under the PT isometry is denoted by $(-\tau, p) \in I^-(\ms{O}) = (-\infty, 0) \times \R^3$.

\medskip

\begin{Def}
\emph{
Consider the PT extension of a Milne-like spacetime. Define the set 
\begin{equation}
I_\ms{O} = I^+(\ms{O})\cup I^-(\ms{O}) \cup \{\ms{O}\} \times \R^3.
\end{equation}
Let $\tau_0 \in \R$. A \emph{cosmic time translation} is a map 
\begin{equation}
f_{\tau_0} \colon I_\ms{O} \to I_\ms{O}.
\end{equation}
If $\tau_0 = 0$, then $f_0$ is just the identity on $I_\ms{O}$. If $\tau_0 \neq 0$, then $f_{\tau_0}$ is given by
\begin{align}
(\tau, p) &\mapsto (\tau + \tau_0, p), \:\:\:\: \text{ for } \tau \neq -\tau_0
\\
 (-\tau_0, p) &\mapsto (\ms{O}, p)
 \\
 (\ms{O},p) &\mapsto (\tau_0, p). 
\end{align}
}
\end{Def}

\medskip

\noindent\emph{Remark.} One can interpret the set $\{\ms{O}\} \times \R^3 \subset I_{\ms{O}}$ as the big bang containing the information of each comoving observer.  

\medskip

 Cosmic time translations take constant $\tau$ slices to other constant $\tau$ slices. Therefore they commute with elements in the Lorentz group. To make this statement precise, for $p \in \R^3$, define $\Lambda p \in \R^3$ via $\Lambda p = q$ where $ \Lambda(\tau, p)=(\tau, q)$. Then for points $(\ms{O}, p) \in \{\ms{O}\} \times \R^3$, define 
\begin{equation} 
 \Lambda (\ms{O}, p) = (\ms{O}, \Lambda p).
 \end{equation}
  Thus $\Lambda$ is now defined on $I_{\ms{O}}$ and $f_\tau \circ \Lambda = \Lambda \circ f_\tau$ for all points in $I_\ms{O}$. The set of all cosmic time translations and $\ms{O}$-fixing causal isometries defined on $I_\ms{O}$ forms a group under composition. 

The elements in this group are isometries on the vector field $\pd_\tau$, so it should be thought of as a \emph{symmetry in cosmic time}. We state this in the theorem below. First note that $f_{\tau_0}$ when restricted to $\big[I^+(\ms{O}) \cup I^-(\ms{O})\big] \setminus \{\tau = -\tau_0\}$ is a map between manifolds, and so its differential, $d(f_{\tau_0})$, is well defined on this restriction.

\medskip

\begin{thm}[Cosmic time invariance]\label{cosmic time invariance}
Consider the \emph{PT} extension $(M,g)$ of a Milne-like spacetime. Let $f_{\tau_0}$ be a cosmic time translation and $\Lambda$ an $\ms{O}$-fixing causal isometry. Then for all $p \in I^+(\ms{O}) \cup I^-(\ms{O})$ with $\tau(p) \neq -\tau_0$, we have
\[
g_p\left( \frac{\pd}{\pd\tau}, \frac{\pd}{\pd\tau}\right) = g_{(f_{\tau_0} \circ \Lambda) p}\left(d(f_{\tau_0} \circ \Lambda)\frac{\pd}{\pd\tau}, d(f_{\tau_0} \circ \Lambda)\frac{\pd}{\pd\tau} \right).
\]

\medskip
\end{thm}

\proof
For $p \in \big[I^+(\ms{O}) \cup I^-(\ms{O})\big]\setminus \{\tau = -\tau_0\}$, the differential $d(f_{\tau_0})_p \colon T_pM \to T_{f_{\tau_0}(p)}M$ makes sense and satisfies 
$
g_p\left( \frac{\pd}{\pd\tau}, \frac{\pd}{\pd\tau}\right) = g_{f_{\tau_0} (p)}\big(d(f_{\tau_0})_p\frac{\pd}{\pd\tau}, d(f_{\tau_0})_p\frac{\pd}{\pd\tau} \big).
$
Therefore the result follows from Theorem \ref{milne isometry}.

\qed

\medskip

Since cosmic time translations and $\ms{O}$-fixing causal isometries commute, this group is isomorphic to $\R \times \text{O}(1,3)$ with isomorphism given by 
\begin{equation}
(\tau_1 + \tau_2, \Lambda_1\Lambda_2) = (\tau_1, \Lambda_1)(\tau_2,\Lambda_2)  \mapsto (f_{\tau_1} \circ \Lambda_1) \circ (f_{\tau_2} \circ \Lambda_2) = f_{\tau_1 + \tau_2} \circ \Lambda_1\Lambda_2.
\end{equation}
Contrast this with the composition law (\ref{poincare composition}) for the Poincar{\'e} group. While the Poincar{\'e} group forms a semi-direct product, the symmetries for a Milne-like spacetime form a direct product. 

\medskip

\begin{Def}
\emph{
 The \emph{symmetry group} for Milne-like spacetimes consists of compositions of $\ms{O}$-fixing causal isometries and cosmic time translations on $I_{\ms{O}}$. It is isomorphic to $\R \times \text{O}(1,3)$. 
}
\end{Def}  

\medskip

The connected component of $\text{O}(1,3)$ containing the identity is the proper orthochronous Lorentz group $\text{L}^\uparrow_+$ (see Appendix \ref{Lorentz group and poincare}). It's double covered by its universal covering group $\text{SL}(2,\C)$ via the homomorphism given by (\ref{cover of proper lorentz}). Therefore $\R \times \text{SL}(2,\C)$ double covers $\R \times \text{L}^\uparrow_+$ and is its universal covering group.

\section{Quantum Theory}


In quantum theory \emph{physical states} are represented by elements in a projective separable complex Hilbert space $\rm{P}\mc{H}$, that is, the set of nonzero vectors $\psi \in \mc{H}$ modulo the equivalence relation $\psi \sim \psi'$ if and only if $\psi = c\psi'$ for some $c \in \C\setminus\{0\}$. We denote the equivalence class by $[\psi] \in \rm{P}\mc{H}$. The inner product $\langle \cdot, \cdot \rangle$ on $\mc{H}$ defines a symmetric real valued function $\langle \cdot, \cdot \rangle \colon \rm{P}\mc{H} \times \rm{P}\mc{H} \to [0,1]$ given by 
\begin{equation}
\langle [\psi], [\phi] \rangle = \frac{|\langle \psi, \phi \rangle |^2}{||\psi||^2||\phi||^2}.
\end{equation}
Physically $\langle [\psi], [\phi] \rangle$ is the \emph{transition probability} of finding the system to be in the state $[\psi]$ when it is in the state $[\phi]$ and vice versa. 

Symmetries are modeled by automorphisms of $\rm{P}\mc{H}$. A bijective map $T \colon \rm{P}\mc{H} \to \rm{P}\mc{H}$ is an \emph{automorphism} of $\rm{P}\mc{H}$ if it preservers the transition probability, i.e. $\langle T[\psi], T[\phi] \rangle = \langle [\psi], [\phi] \rangle$ for all $[\psi],[\phi] \in \rm{P}\mc{H}$. If $U$ is a unitary or anti-unitary operator on $\mc{H}$, then it defines an automorphism $[U]$ on $\rm{P}\mc{H}$ by $[U][\psi] = [U\psi]$. The product of two anti-unitary operators is unitary, so the group $\rm{U}(\mc{H})$ of unitary operators is a subgroup of index two of the group $\tilde{\rm{U}}(\mc{H})$ of unitary or anti-unitary operators. 

The map $e^{i\theta} \mapsto e^{i\theta}\cdot 1_{\mc{H}}$ embeds $\rm{U}(1)$ as a subgroup of $\rm{U}(\mc{H})$. Let $\rm{Aut}(\rm{P}\mc{H})$ be the group of automorphisms of $\rm{P}\mc{H}$ and $\pi \colon \tilde{\rm{U}}(\mc{H}) \to \rm{Aut}(\rm{P}\mc{H})$ be the map $U \mapsto [U]$. Wigner showed that $\pi$ is actually a surjection, that is, any automorphism on $\rm{P}\mc{H}$ is of the form $[U]$ for some unitary or antiunitary $U \in \tilde{\rm{U}}(\mc{H})$. For a proof, see e.g. \cite[Ch. 2 Appendix A]{Weinberg}. Specifically we have

\medskip

\begin{thm}[Wigner]\label{wigner thm}
The sequence 
\begin{equation}\label{wigner}
\xymatrix
{
1\ar[r] &\rm{U}(1)\ar[r] &\tilde{\rm{U}}(\mc{H}) \ar[r]^{\pi\:\:\:\:\:\:} &\rm{Aut}(\rm{P}\mc{H}) \ar[r] &1
}
\end{equation}
is exact. 
\end{thm}

\medskip

 Let $\rm{U}(\rm{P}\mc{H})$ be the image of $\rm{U}(\mc{H})$ under $\pi$. If $G$ is a connected Lie group representing the group of symmetries on a spacetime manifold, then the image of any homomorphism $T\colon G \to \rm{Aut}(\rm{P}\mc{H})$ is contained in $\rm{U}(\rm{P}\mc{H})$. To see this note that in a neighborhood $W$ of the identity $1_G$, the exponential map allows one to write each element of $W$ as a square of another element in $W$. This combined with the fact that $W$ generates $G$ \cite[Proposition 7.14]{Lee} proves the claim. In this case the exact sequence (\ref{wigner}) restricts to the exact sequence
\begin{equation}
\xymatrix
{
1\ar[r] &\rm{U}(1)\ar[r] &\rm{U}(\mc{H}) \ar[r]^{\pi\:\:\:\:} &\rm{U}(\rm{P}\mc{H}) \ar[r] &1
}
\end{equation}
 
  We endow $\rm{U}(\rm{P}\mc{H})$ with the weakest topology such that all maps $\rm{U}(\rm{P}\mc{H}) \to \rm{P}\mc{H}$, $[U] \mapsto [U][\psi] = [U\psi]$, are continuous for each $[\psi] \in \rm{P}\mc{H}$. We endow $\rm{U}(\mc{H})$ with the strong operator topology, i.e. the weakest topology such that all maps $\rm{U}(\mc{H}) \to \mc{H}$, $U \mapsto U \psi$, are continuous for each $\psi \in \mc{H}$. A \emph{projective unitary representation} of $G$ is a continuous homomorphism $r \colon G \to \rm{U}(\rm{P}\mc{H})$. A \emph{unitary representation} of $G$ is a continuous homomorphism $R \colon G \to \rm{U}(\mc{H})$. (We will sometimes omit the word `unitary' for brevity). Clearly $\pi \circ R$ is a projective unitary representation whenever $R$ is a unitary representation. Conversely, if $r \colon G \to \rm{U}(\rm{P}\mc{H})$ is a projective unitary representation, then $r$ \emph{admits a lifting $R$} if there exists a unitary representation $R \colon G \to \rm{U}(\mc{H})$ such that $r = \pi \circ R$. 
  
%
%
%
  If every projective unitary representation lifts, then we can classify the automorphisms in $\rm{U}(\rm{P}\mc{H})$ (i.e. the symmetries) by unitary representations. A theorem of Bargmann \cite{Barg1} shows such liftings occur when the second cohomology group of the Lie algebra $\mathfrak{g}$ of a connected and simply connected Lie group $G$ is trivial.
  
\medskip

\begin{thm}[Bargmann]\label{Bargmann Theorem}
Let $G$ be a connected and simply connected Lie group. If $H^2(\mathfrak{g}, \R) = 0$, then any projective unitary representation lifts to a unitary representation. 
\end{thm}

\medskip

Let $R\colon G \to \rm{U}(\mc{H})$ be a unitary representation. A closed subspace $S \subset \mc{H}$ is an \emph{invariant subspace} for $R$ if $R(g)S \subset S$ for all $g \in G$. If the only invariant subspaces of $R$ are $\{0\}$ and $\mc{H}$, then $R$ is \emph{irreducible}. An \emph{elementary particle} is an irreducible unitary representation $R \colon G \to \rm{U}(\mc{H})$. The physical motivation is that if there was a Hilbert space $\mc{H}$ containing our elementary particle, then we would expect there is no nontrivial subspace $S$ in which vectors can be simply transformed by $R$, otherwise $S$ would be `more elementary'. 

In the next section we determine the irreducible unitary representations for the Milne-like symmetry group and compare and contrast this with the irreducible unitary representations of the Poincar{\'e} group.

\medskip

\subsection{Irreducible Unitary Representations of the Milne-like Symmetry Group}\label{irrep for milne}

Consider the PT extension of a Milne-like spacetime. We postulate that physical quantities do not depend on the symmetry group $\R \times \text{O}(1,3)$. Therefore these quantities should be invariant under compositions of $\ms{O}$-fixing causal isometries and cosmic time translations. We will see that these quantities correspond to the mass and spin of particles. Since we only want to work with unitary operators, we consider only those transformations which can be linked back to the identity. Hence we work with $\R \times \text{L}^\uparrow_+$ which is the connected component of $\R \times \text{O}(1,3)$ which contains the identity. We form the following postulate in analogy to the relativistic invariance postulate.

\medskip

\noindent{\bf Cosmological Invariance Postulate.} \emph{There is a projective unitary representation of $\R \times \emph{L}^\uparrow_+$ into $\emph{\text{U}}(\emph{\text{P}}\mc{H})$ where $\mc{H}$ is the Hilbert space of physical states.}

\medskip

We have $H^2(\mathfrak{g}, \R) = 0$ where $\mathfrak{g} = \text{Lie}(\R \times \text{L}^\uparrow_+) \approx \R \oplus \mf{so}(1,3)$ with Lie bracket given by equation (\ref{milne lie bracket}). This follows from the K{\"u}nneth formulas \cite[Ch. XI Theorem 3.1]{CartanEilenberg}. Alternatively, one can show this directly following an approach similar to \cite[Section 2.7]{Weinberg}.  Bargmann's Theorem \ref{Bargmann Theorem}  applies. Thus any projective unitary representation of the universal covering group $\R \times \rm{SL}(2,\C)$ of $\R \times \text{L}^\uparrow_+$ lifts to a unitary representation.

To determine the elementary particles, we want to classify the irreducible unitary representations of $\R \times \rm{SL}(2, \C)$.

\medskip

\subsubsection{The Classification}

The irreducible unitary representations of $\R \times \text{SL}(2, \C)$ are determined by the irreducible unitary representations of $\text{SL}(2,\C)$. More precisely,

\medskip

\begin{thm}\label{irreps for Milne}
A unitary representation $R \colon \R \times \emph{\text{SL}}(2,\C) \to \emph{\text{U}}(\mathcal{H})$ is irreducible if and only if $R(\tau, h) = e^{-im\tau}\s(h)$ for some $m \in \R$ and some irreducible unitary representation  $\s \colon \emph{\text{SL}}(2,\C) \to \emph{\text{U}}(\mathcal{H})$. 
\end{thm}

\proof
Suppose $\s \colon \text{SL}(2,\C) \to \text{U}(\mathcal{H})$ is an irreducible unitary representation. Fix $m \in \R$ and define $R(\tau, h) = e^{-im\tau}\s(h)$. Then $R\colon \R \times \text{SL}(2,\C) \to \text{U}(\mathcal{H})$ is a unitary representation. Let $S \subset \mathcal{H}$ be a nonzero invariant subspace of $R$. Suppose $S \neq \mathcal{H}$. Then there exists a $\psi \in \mathcal{H} \setminus S$ such that $R(\tau,h)\psi \notin S$ for all $(\tau, h)$. But this implies $\s(h)\psi \notin S$ for all $h$. This contradicts $\s$ being irreducible. Thus $S = \mathcal{H}$, and so $R$ is an irreducible unitary representation. 

Conversely, suppose $R\colon \R \times \text{SL}(2,\C) \to \text{U}(\mathcal{H})$ is an irreducible unitary representation. Define $\xi(\tau) = R(\tau, 1)$. Since $R$ is a homomorphism, we have $\xi(\tau)R(\tau',h) = R(\tau',h)\xi(\tau)$.
By Schur's lemma \cite[Theorem 3.5 (a)] {Folland}, we find $\xi(\tau)$ is a constant multiple of the identity. This constant must have norm 1 because $R$ is unitary. Therefore there exists an $m \in \R$ such that $\xi(\tau) = e^{-im\tau}\cdot 1_{\mathcal{H}}$. Define $\s(h) = R(0, h)$. Then $R(\tau,h) = e^{-im\tau}\s(h)$. The same argument above shows that if $S \subset \mathcal{H}$ is an invariant subspace for $R$, then $S$ is an invariant subspace for $\s$. Thus $\s$ is irreducible.

\qed

\medskip

Let $G = \R \times \text{SL}(2,\C)$. For each $m \in \R$, the \emph{orbit} of $m$ is the set $O_m = \{g^{-1}(m,1)g \mid g \in G\}$. Therefore $O_m = \{(m,1)\}$. The \emph{stabilizer} of $m$ is $G_m = \{g \in G \mid g^{-1}(m,1)g = (m,1)\}$. Therefore $G_m \approx \text{SL}(2,\C)$. The \emph{little group} of $m$ is the set $H_m = G_m \cap \text{SL}(2,\C)$. Therefore $H_m \approx \text{SL}(2,\C)$. In general the irreducible unitary representations for representative points on an orbit are determined by the irreducible unitary representations of its little group. This is Mackey's Theorem \cite[Theorem 6.42]{Folland}. Theorem \ref{irreps for Milne} above is a special case of this.

Let $R(\tau,h) = e^{-im\tau}\s(h)$ be an irreducible unitary representation of $\R \times \text{SL}(2,\C)$. Then $R(\tau,1) = e^{-im\tau}\cdot 1_\mathcal{H}$.  Let $M$ be the mass operator from Appendix \ref{casimir for milne-like}. From (\ref{def of self-adjoint R}) 
\begin{equation}
\exp(-i\tau M) = R\big(\exp(\tau \pi),1 \big) = R(\tau,1) = e^{-im\tau}\cdot 1_{\mc{H}}.  
\end{equation}
Therefore $M = m\cdot 1_{\mc{H}}$ on each orbit $O_m$ by Schur's lemma. Since $M$ is a Casimir operator, the mass $m$ of an orbit is an observer-independent quantity. The masses and little groups for each of the orbits are

\medskip

\begin{center}
 \begin{tabular}{c| c |c} 
 
 Orbit $O_m$ & Representative point $m$ & Little Group $H_m$   \\ [.75ex] 
 \hline
 & & 
 \\
 $O_m^+$ & $|m|$ & $\text{SL}(2,\C)$ 
  \\ [.75ex] 

 $O_m^-$ & $-|m|$ & $\text{SL}(2,\C)$ 
  \\[.75ex] 

 $O_0$ & $0$ & $\text{SL}(2,\C)$ 
 \\ [.75ex] 

\end{tabular}
\end{center}

\medskip

\newpage

\noindent{\bf A Physical Interpretation}:
\\
 The orbit $O^m_+$ corresponds to particles with positive mass $|m| >0$ traveling along increasing $\tau$. They comprise the comoving observers in $I^+(\ms{O})$. The orbit $O^m_-$ corresponds to particles with negative mass $-|m| < 0$ traveling along increasing $\tau$. Therefore they have mass $|m| > 0$ traveling along \emph{decreasing} $\tau$. They comprise the comoving observers in $I^-(\ms{O})$. The orbit $O^0$ consists of massless particles. They correspond to particles traveling along $\tau = 0$. Equation (\ref{def for tau}) implies these are null curves. Therefore massless particles travel at the speed of light. Each distinctive orbit $O^+_m,$  $O^-_m,$ and $O_0$ has a physical interpretation. This is unlike the Poincar{\'e} group where the majority of the orbits lack any real physical interpretation. 

\medskip

To finish the classification we must determine the irreducible unitary representations of $\text{SL}(2,\C)$. This was done by Bargmann \cite{Barg2}. The other Casimir operators from Appendix \ref{casimir for milne-like} are $Q = \delta_{ij}(J^iJ^j - K^iK^j)$ and $S = \delta_{ij}K^iK^j$. From \cite{Barg2} the spectrum of $\delta_{ij}J^iJ^j$ is discrete and consists of the form $j(j+1)$ where $j$ is the spin of the particle and can take on values $0,\: 1/2,\: 1,\: 3/2,\dotsc$ These correspond to the $(2j+1)$-dimensional irreducible representations of $\text{SO}(3)$, every one of which occurs exactly once. Given an irreducible representation, Schur's lemma implies $Q = q\cdot 1_{\mc{H}}$ and $S = s\cdot 1_{\mc{H}}$ on each orbit. The irreducible representations may be classified into two groups: (i) $q > 0$, $s = 0$, and $j$ assumes all values $0,\: 1,\: 2, \dotsc$ (ii) $s$ can be any real number and $q = 1 - k^2 + (s/k)^2$ where $k$ may have anyone of the values $1/2,\: 1,\: 3/2, \dotsc$ and $j$ assumes all values $k$, $k+1$, $k+2, \dotsc$ 

The symmetry group for Milne-like spacetimes has three Casimir operators, while the symmetry group for Minkowski space has two Casimir operators. It would be interesting if there are any physical consequences of this distinction. For example, half-integer spin particles (e.g. electrons) must belong to group (ii) and any spin 0 particle (e.g. the Higgs boson) must belong to group (i).

\medskip\medskip

\subsubsection{Contrasting with the Poincar{\'e} Group}

Traditionally, the Poincar{\'e} group, $\R^4 \rtimes \text{O}(1,3)$, is used as the symmetry group for quantum theory. This is stated in the following postulate.

\medskip

\noindent{\bf Relativistic Invariance Postulate.} \emph{There is a projective unitary representation of 
\linebreak
$\R^4 \rtimes \emph{L}^\uparrow_+$ into $\emph{\text{U}}(\emph{\text{P}}\mc{H})$ where $\mc{H}$ is the Hilbert space of physical states.}

\medskip

Bargmann's Theorem applies, and so any projective unitary representation of the universal covering group $\R^4 \rtimes \text{SL}(2,\C)$ of $\R^4 \rtimes \text{L}^\uparrow_+$ can be lifted to a unitary representation. The orbits for $\R^4$ under the $\text{SL}(2,\C)$ action are

\begin{align*}
O^m_+ &= \{p \in \R^4 \mid -p_\mu p^\mu = m^2 >0, \:\: p^0 > 0\},
\\
O^m_- &= \{p \in \R^4 \mid -p_\mu p^\mu = m^2 >0, \:\: p^0 < 0\},
\\
O^0_+ &= \{p \in \R^4 \mid -p_\mu p^\mu = 0, \:\: p^0 > 0\},
\\
O^0_- &= \{p \in \R^4 \mid -p_\mu p^\mu = 0, \:\: p^0 < 0\},
\\
T^{m} &= \{p \in \R^4 \mid -p_\mu p^\mu = -m^2 > 0\},
\\
\{0\}.
\end{align*}

The quantity $-p_\mu p^\mu$ corresponds to the eigenvalue of the mass-squared operator, $M^2 = -P^\mu P_\mu$,  given in Appendix \ref{casimir for poincare}. By Mackey's Theorem, the irreducible unitary representations of the Poincar{\'e} group are determined by the irreducible unitary representations of the following little groups.

\begin{center}
 \begin{tabular}{c| c |c | c} 
 
 Orbit $O_p$ & Representative point $p$ & Little Group $H_p$  & Mass \\ [.75ex] 
 \hline
 & & & 
 \\
 $O^m_+$ & $(|m|,0,0,0)$ & $\rm{SU}(2)$ & $|m|$ 
  \\ [.75ex] 

 $O^m_-$ & $(-|m|,0,0,0)$ & $\rm{SU}(2)$ & $|m|$
  \\[.75ex] 

 $O^0_+$ & $(1,1,0,0)$ & $\Delta$ &0
 \\[.75ex] 
 
 $O^0_-$ & $(-1,1,0,0)$ & $\Delta$ &0
 \\ [.75ex] 
 
 $T^m$ & $(0,|m|,0,0)$ & $\rm{SL}(2,\R)$ &$i|m|$
 
  \\ [.75ex] 
  
 $\{0\}$ & $(0,0,0,0)$ & $\rm{SL}(2,\C)$ & $0$ 
 \\ [.75ex] 

\end{tabular}
\end{center}
where 
\begin{equation}
\Delta  = \left\{ 
\left(\begin{array}{cc} e^{i\theta} & z\\ 0 & e^{-i\theta} \end{array}\right) \mid \theta \in \R, \:\: z \in \C \right\}.
\end{equation}

\noindent{\bf The Physical Interpretation}:
\\
 $O^m_+$ and $O^0_+$ are the \emph{only} orbits which correspond to known particles, and so these are the only orbits which have a physical interpretation. This is unlike the classification for the Milne-like symmetry group were each orbit has a physical interpretation. 

$O^m_+$ corresponds to particles with mass $m >0$ and positive energy. $O^0_+$ corresponds to particles with zero mass and positive energy. The orbits $O^m_-$ and $O^0_-$ correspond to particles with negative energy which have never been observed. The orbit $T^m$ corresponds to faster-than-light particles. These have obviously never been observed. The orbit $\{0\}$ is sometimes said to ``represent the vacuum" \cite{FollandQFT}. If so, then what is the significance of the little group $\text{SL}(2,\C)$ for the vacuum $\{0\}$?

Let $W^\mu W_\mu$ be the other Casimr operator from Appendix \ref{casimir for poincare}. Choosing $p = (|m|, 0 , 0 , 0)$ as a representative point for $O^m_+$, we have $W^\mu W_\mu = m^2 \delta_{ij}J^iJ^j$. The operator $\delta_{ij}J^iJ^j$ is Casimir for the little group $\text{SU}(2)$ with eigenvalues $j(j+1)$ where $j$ is the spin of the particle and can take on values $0, \: 1/2, \: 1, \: 3/2, \dotsc$. The irreducible representations of $\text{SU}(2)$ are well understood \cite{Folland, tung} and are characterized by $j$.  As explained in \cite{Folland, tung, Simms}, for the orbit $O^0_+$ there are two families of irreducible representations for $\Delta$ . The first family is discrete and corresponds to values $\theta = n \in \Z$ and $z = 0$. The other family is continuous and corresponds to values $z \in \C \setminus \{0\}$. No known physical particles have been observed from the continuous family.

\subsection{The Dirac Equation on Milne-like Spacetimes}

\subsubsection{Motivating the Equation}

Consider the PT extension $(M,g)$ of a Milne-like spacetime. By a \emph{spinor field} $\psi$ we mean a smooth function on $M$ into spinor space. Since $M \subset \R^4$ can be chosen to be simply connected (see Figure \ref{matter-antimatter}), this definition makes sense \cite[Chapter 13]{Wald}. We emphasize that our spinor fields are not operators, so this is the step before `second quantization.' The mass operator from section \ref{irrep for milne} is a Casimir operator generated from cosmic time translations, so we postulate that the mass operator corresponds to $i \pd_\tau$ and that every spinor field satisfies
\begin{equation}\label{mass eigen}
i \pd_\tau \psi = m\psi. 
\end{equation}
Recall that the vector field $\pd_\tau$ is defined for all points $I^+(\ms{O}) \cup I^-(\ms{O})$ and is always future pointing (even in $I^-(\ms{O})$). It's the vector field associated with the comoving observers. 
There is nothing in equation (\ref{mass eigen}) which forbids $m$ from being negative. If $m > 0$, then (\ref{mass eigen}) can be interpreted as an observer measuring an energy $m$ in the particle's rest frame, i.e. $E = mc^2$. If $m < 0$, then (\ref{mass eigen}) may be interpreted as the energy $E = |m|$ measured by an observer moving in the $-\pd_\tau$ direction. Equation (\ref{mass eigen}) is manifestly Lorentz invariant.

Now we make the connection to the Dirac equation. Let $x^\mu = (t, x, y, z)$ be a coordinate system for the PT extension. In these coordinates the metric takes the form $g = \Omega^2(\tau)\eta$ where $\eta$ is the usual Minkowski metric. With respect to these coordinates, an observer will measure an energy $p^0$ and momentum $p^i$ of a particle with mass $m$ such that $
 -g_{\mu\nu}p^\mu p^\nu = -\Omega^2(\tau)\eta_{\mu\nu}p^\mu p^\nu = m^2.
$
We assume the usual substitution $p^\mu \to i\pd^\mu$. For a spin 0 field $\psi$, this yields the \emph{Klein-Gordon equation} for Milne-like spacetimes 
$
\big[\Omega^2(\tau)\eta^{\mu\nu}\pd_\mu\pd_\nu\big] \psi = m^2 \psi.
$
Likewise, for a Dirac spinor $\psi$, we have the \emph{Dirac equation}\footnote{One could also arrive at this equation from the Dirac equation on curved spacetimes. See \cite[Ch. 13]{Wald}.} for Milne-like spacetimes  

\begin{equation}\label{dirac for milne}
\big[\Omega(\tau)\gamma^\mu\pd_\mu\big] \psi = m\psi.
\end{equation}

Recall we are using the $(-, +, +, +)$ signature convention. Since the Dirac equation is just the product of $\Omega$ and the original Dirac equation, many of the properties carry over. For example, there is still a conserved probability current. Also, our Dirac equation is Lorentz invariant because the original Dirac equation is Lorentz invariant and $\Omega(\Lambda \tau) = \Omega(\tau)$ for any $\ms{O}$-fixing causal isometry $\Lambda \in \text{O}(1,3)$.

When solving the Dirac equation, we distinguish between two separate scenarios.

\medskip

\begin{Def}\label{solving dirac}
\emph{
\begin{itemize}
\item We say $\psi$ \emph{solves the Dirac equation for $I^+(\ms{O})$} if 
\begin{equation}
\big[\Omega(\tau)\gamma^\mu\pd_\mu\big] \psi = m\psi.
\end{equation}
\item We say $\psi$ \emph{solves the Dirac equation for $I^-(\ms{O})$} if
\begin{equation}\label{dirac for PT milne}
\big[ \Omega(\tau)\g^\mu (-\pd_\mu)\big] \psi = m\psi.
\end{equation}
\end{itemize}
}
\end{Def}

\medskip

The idea behind Definition \ref{solving dirac} is that the observers in $I^+(\ms{O})$ would use $x^\mu$ as their coordinates while observers in $I^-(\ms{O})$ would use $-x^\mu$ as their coordinates. Note that equation (\ref{dirac for PT milne}) is equivalent to $\big[\Omega(\tau)\gamma^\mu\pd_\mu\big] \psi = -m\psi$. Whether $\psi$ solves the Dirac equation for $I^+(\ms{O})$ or $I^-(\ms{O})$, the anticommutation Clifford relations imply 
\begin{equation}
\Omega^2(\g^\mu\pd_\mu)(\g^\nu\pd_\nu)\psi = \Omega^2\big(\g^\mu(-\pd_\mu)\big)\big(\g^\nu(-\pd_\nu)\big) \psi = m^2\psi.
\end{equation}
In the \emph{Weyl representation}, the matrices $\g^\mu$ are
\begin{equation}
\g^0= 
i\left(\begin{array}{cc} 0 & I\\ I & 0 \end{array}\right), \:\:\:\: \g^j = i\left(\begin{array}{cc} 0 & \s^j\\ -\s^j & 0 \end{array}\right).
\end{equation}
We use the Weyl representation because we can take take advantage of  (\ref{vector to matrix}).
Choose a single comoving observer in the PT extension. Pick coordinates $(t,x,y,z)$ so that they are aligned with this comoving observer. Then $x = y = z =0$ along the observer's timelike geodesic. From equation (\ref{t and r}), the relationship between $\tau$ and $t$ is $b(\tau) = t$. Therefore $\pd_t = b'(\tau)\pd_\tau$. Hence $\pd_\tau = \Omega(\tau)\pd_t$, and so equations (\ref{mass eigen}) and (\ref{dirac for milne}) agree. This gives more credence that cosmic time translations should be part of the symmetry group for quantum theory.

\subsubsection{PT Symmetry}

The PT symmetry of the original Dirac equation is well known. Therefore the corresponding results for the Dirac equation on Milne-like spacetimes carry over easily. The crucial difference for Milne-like spacetimes is the \emph{context}. We believe the PT symmetry of the Dirac equation supports the claim that the universe's missing antimatter comprises $I^-(\ms{O})$. In this section we quickly review the PT symmetry of the Dirac equation.

We introduce electromagnetism in the Dirac equation. Assume there is an electromagnetic field tensor $F_{\mu\nu}$ on the PT extension $M$. Since we can choose $M$ to be simply connected (see Figure \ref{matter-antimatter}), the converse of the Poincar{\'e} lemma ensures there is a one-form electromagnetic potential $A_\mu$ on $M$ such that$F_{\mu\nu} = \pd_\mu A_\nu - \pd_\nu A_\mu$. We assume the usual prescription $i\pd_\mu \to i\pd_\mu - eA_\mu$, or equivalently, $\pd_\mu \to \pd_\mu + ieA_\mu$. Here $e$ can be any charge. We extend Definition \ref{solving dirac} to include $A_\mu$.

\begin{Def}
\emph{
\begin{itemize}
\item We say $\psi$ \emph{solves the Dirac equation for $I^+(\ms{O})$ with potential $A_\mu$}, if 
\begin{equation}\label{dirac solve with pot}
\Omega \g^\mu(\pd_\mu + i e A_\mu) \psi = m \psi
\end{equation}
\item We say $\psi$ \emph{solves the Dirac equation for $I^-(\ms{O})$ with potential $A_\mu$}, if
\begin{equation}
\Omega \g^\mu(-\pd_\mu + i e A_\mu) \psi = m \psi
\end{equation}
\end{itemize}
}
\end{Def}

\medskip

\begin{prop}\label{star}
If $\psi$ solves the Dirac equation for $I^+(\ms{O})$ with potential $A_\mu$, then the complex conjugate $\psi^*$ solves the Dirac equation for $I^-(\ms{O})$ with potential $A_\mu$.
\end{prop}

\proof
Taking the complex conjugate of equation (\ref{dirac solve with pot}) yields
\begin{align*}
m\psi^* &= \Omega(\g^\mu)^*(\pd_\mu - ieA_\mu)\psi^* 
\\
&= \Omega(-\g^\mu)(\pd_\mu - ieA_\mu)\psi^*
\\
&=\Omega\g^\mu(-\pd_\mu + ieA_\mu)\psi^* 
\end{align*}
\qed

%

\medskip

\noindent Define 
\begin{equation}
\g(x) = \sum_{\mu =0}^3 x^\mu\g^\mu =
i\left(\begin{array}{cc} 0 & \underline{x}\\ \underline{\text{P}x} & 0 \end{array}\right) 
\end{equation}
where $\underline{x}$ is given by equation (\ref{vector to matrix}) and $\text{P}x = (x^0, -x^1, -x^2, -x^3)$. Let $\text{PT} \in \text{GL}(4,\C)$ be an element which reverses both space and time. There are two choices which differ by a negative sign. Let's choose
\begin{equation}
\text{PT} = \left(\begin{array}{cc} I & 0 \\ 0 & -I \end{array}\right).
\end{equation}
Then $\g(-x) = \text{PT}\; \g(x)(\text{PT})^{-1}$. Hence $\text{PT}$ reverses space and time by acting on $\g(x)$ via conjugation. Note that $ \text{PT}\;\g^\mu = - \g^\mu\;\text{PT} .$

\medskip

\begin{prop}\label{PT psi}
If $\psi$ solves the Dirac equation for $I^+(\ms{O})$ with potential $A_\mu$, then $\emph{\text{PT}}\psi$ solves the Dirac equation for $I^-(\ms{O})$ with potential $-A_\mu$. 
\end{prop}

\proof

\begin{align*}
 \Omega\g^\mu\big[-\pd_\mu -ieA_\mu\big]\text{PT} \psi
&= \Omega(\g^\mu \text{PT})\big[-\pd_\mu -ieA_\mu\big] \psi
\\
&= \Omega(-\text{PT}\g^\mu)\big[-\pd_\mu -ieA_\mu\big] \psi
\\
&=\text{PT}\;\Omega\g^\mu\big[\pd_\mu + ieA_\mu\big]\psi
\\
&=m\text{PT} \psi.
\end{align*}
\qed

%

\medskip

\noindent Taking the complex conjugate of the result from Proposition \ref{PT psi} gives

\begin{prop}\label{PT psi star}
If $\psi$ solves the Dirac equation for $I^+(\ms{O})$ with potential $A_\mu$, then $ \emph{\text{PT}}\psi^*$ solves the Dirac equation for $I^+(\ms{O})$ with potential $-A_\mu$.
\end{prop}

%

\medskip

 \noindent We summarize Propositions \ref{star}, \ref{PT psi}, and \ref{PT psi star} in the following table.
 
 \medskip

\begin{center}
 \begin{tabular}{c| c |c} 
 
 Spinor field & Equation & An interpretation   \\ [.75ex] 
 \hline
 & & 
 \\
 $\psi$ & $\Omega \g^\mu(\pd_\mu + ieA_\mu)\psi = m\psi$ & $\psi$ in $I^+(\ms{O})$
  \\ [.75ex] 

 $\psi^*$ & $\Omega \g^\mu(-\pd_\mu + ieA_\mu)\psi^* = m\psi^*$ & $\psi$ in $I^-(\ms{O})$ 
  \\[.75ex] 

 $\text{PT}\psi$ & $\Omega \g^\mu(-\pd_\mu - ieA_\mu)\text{PT}\psi = m\text{PT}\psi$ & Anti $\psi$ in $I^-(\ms{O})$  
 \\ [.75ex] 
 
 $\text{PT}\psi^*$ & \:\: $\Omega \g^\mu(\pd_\mu - ieA_\mu)\text{PT}\psi^* = m\text{PT}\psi^*$ \:\: & Anti $\psi$ in $I^+(\ms{O})$ 
 \\ [.75ex] 

\end{tabular}
\end{center}

\medskip

Given the interpretation, we believe the relationship between $\psi$ and $\text{PT}\psi$  suggests that our universe's missing antimatter comprises $I^-(\ms{O})$.

\subsection*{Acknowledgments}
We thank Greg Galloway, Nikolai Saveliev, and Alexei Deriglazov for helpful comments and discussions.

\appendix

\section{The Poincar{\'e} Group and its Double Cover}\label{Lorentz group and poincare}

Let $(\R^4, \eta)$ be Minkowksi space. The isometries on Minkowski space are of the form
\begin{equation}
x'^\mu = \Lambda^\mu_{\:\:\:\nu}x^\nu + a^\nu
\end{equation}
where $\Lambda^\mu_{\:\:\:\nu}$ is an element of the Lorentz group $\text{L} = \rm{O}(1,3)=\{\Lambda \mid \Lambda^\a_{\:\:\:\mu}\Lambda^\b_{\:\:\:\nu}\eta_{\a\b} = \eta_{\mu\nu}\}$ and $a^\mu$ is a four-vector. The transformation from $x$ to $x'$ is a Lorentz rotation $\Lambda$ followed by a spacetime translation $a$. Transformation of two isometries yields
\begin{equation}
x''^\mu = \bar{\Lambda}^\mu_{\:\:\:\nu}x'^\nu + \bar{a}^\mu =  \bar{\Lambda}^\mu_{\:\:\:\nu}\Lambda^\nu_{\:\:\:\a}x^\a + (\bar{\Lambda}^\mu_{\:\:\:\nu}a^\nu + \bar{a}^\mu). 
\end{equation}
Therefore the set of all such pairs $(a,\Lambda)$ forms a group with composition law 
\begin{equation}\label{poincare composition}
(a_1, \Lambda_1)(a_2,\Lambda_2) = (a_1 + \Lambda_1a_2, \Lambda_1\Lambda_2).
\end{equation}
This is the \emph{Poincar{\'e}} group $\ms{P} = \R^4 \rtimes \text{O}(1,3)$ which is the \emph{symmetry group} for Minkowski space. The Lorentz group $\text{L} = \text{O}(1,3)$ has four connected components $\text{L}^\uparrow_{+}$, $\text{L}^\downarrow_+$, $\text{L}^\uparrow_{-}$, $\text{L}^\downarrow_{-}$. The $\pm$ corresponds to $\det \Lambda = \pm 1$, the $\uparrow$ corresponds to $\Lambda^0_{\:\:\:0} \geq 1$, and the $\downarrow$ corresponds to $\Lambda^0_{\:\:\:0} \leq -1$. The \emph{proper orthochronous Lorentz group} is $\text{L}^\uparrow_{+}$; it's the connected component which contains the identity. The connected components of the Lorentz group  divide the Poincar{\'e} group $\ms{P}$ into four corresponding connected components $\ms{P}^\uparrow_+$, $\ms{P}^\downarrow_{+}$, $\ms{P}^\uparrow_{-}$, $\ms{P}^\downarrow_-$. The subgroup $\ms{P}^\uparrow_+$ contains the identity. It is the \emph{restricted Poincar{\'e} group}.

 Define a surjective homomorphism 
\begin{equation}\label{cover of proper lorentz} 
 \rm{SL}(2,\C) \to L^\uparrow_+ \:\:\:\: \text{ by } \:\:\:\: \Lambda^\mu_{\:\:\:\nu}(A) = \frac{1}{2}\rm{Tr}\big[\sigma^\mu A\sigma^\nu A^\dagger \big] 
\end{equation} 
 where $\sigma^0, \sigma^1, \sigma^2, \sigma^3$ are the usual Pauli matrices. 
 
 \begin{equation}
\s^0= 
\left(\begin{array}{cc} 1 & 0\\ 0 & 1 \end{array}\right) \:\:\:\: 
\s^1= 
\left(\begin{array}{cc} 0 & 1\\ 1 & 0 \end{array}\right)\:\:\:\: 
\s^2= 
\left(\begin{array}{cc} 0 & -i\\ i & 0 \end{array}\right)\:\:\:\: 
\s^3= 
\left(\begin{array}{cc} 1 & 0\\ 0 & -1 \end{array}\right)
\end{equation}
 
 One readily checks that $\Lambda(A_1A_2) = \Lambda(A_1)\Lambda(A_2)$. Moreover $\Lambda(A_1) = \Lambda(A_2)$ if and only if $A_1 = \pm A_2$. Hence $\rm{SL}(2,\C) \to L^\uparrow_+$ is a double cover. Given a four-vector $x = (x^0,x^1,x^2,x^3)$, we define a Hermitian matrix $\underline{x} = -x^\mu\sigma_\mu = -x^\mu\sigma^\nu\eta_{\mu\nu}$.
  \begin{equation}\label{vector to matrix}
\underline{x}= 
\left(\begin{array}{cc} x^0 + x^3 & x^1 - ix^2\\ x^1 + ix^2 & x^0 - x^3 \end{array}\right)  
\end{equation}
 
  We recover $x$ from $\underline{x}$ by $x^\mu = \frac{1}{2}\rm{Tr}(\sigma^\mu \underline{x})$. If $x'^\mu = \Lambda^\mu_{\:\:\:\nu}(A)x^\nu$, then $\underline{x}' = A \underline{x}A^\dagger$. $\rm{SL}(2,\C)$ is topologically $S^3 \times \R^3$ so it's simply connected; hence it's the universal covering group of $\text{L}^\uparrow_{+}$. The \emph{restricted spinor group} $\ms{P}_0$ consists of all pairs ($\underline{a}, A)$ with $\underline{a}$ Hermitian and $A \in \rm{SL}(2,\C)$. The group operation is
\begin{equation}
(\underline{a}_1, A_1)(\underline{a}_2,A_2) = (\underline{a}_1 + A_1\underline{a}_2A_1^\dagger, A_1A_2).
\end{equation}
 Therefore it's isomorphic to the semi-direct product $\ms{P}_0 \approx \R^4 \rtimes \rm{SL}(2,\C)$. The element $(\underline{a}, A) \in \ms{P}_0$ corresponds to the Poincar{\'e} transformation $\underline{x}' = A\underline{x}A^\dagger + \underline{a}$. The double cover homomorphism $\rm{SL}(2, \C) \to L^\uparrow_{+}$ induces a double cover homomorphism $\ms{P}_0 \to \ms{P}^\uparrow_+$. Thus $\ms{P}_0$ is the universal covering group of the restricted Poincar{\'e} group $\ms{P}^\uparrow_+$.

The Lie algebra of $\R^4$ is identified with itself with trivial Lie bracket. The Lie algebra of $\rm{SL}(2,\C)$ is isomorphic to the Lie algebra of $\text{SO}(1,3)$ which is the set of $4\times 4$ matrices which are skew relative to $\eta$
\begin{equation}
\mf{so}(1,3) = \{A \mid \eta_{\a\mu}A^{\b\mu} = -A^{\a\mu}\eta_{\mu\b}\} = \{A \mid A^\b_{\:\:\:\a} = - A^{\a}_{\:\:\:\b}\}.
\end{equation}
Hence a typical element $A \in \mf{so}(1,3)$ is of the form 
\begin{equation}
A = \left(\begin{array}{cccc} 
0 & a & b & c
\\
a & 0 & d & e
\\
b & -d & 0 & f
\\
c & -e & -f & 0
\end{array}\right),
\:\:\:\: \text{ with } a,b,c,d,e,f\in \R. 
\end{equation}
Therefore the Lie algebra of $\R^4 \rtimes \rm{SL}(2,\C)$ is the vector space $\R^4 \oplus \mf{so}(1,3)$ with Lie bracket 
\begin{equation}\label{lie algebra for poincare}
\big[(x,A), (y,B) \big] = \big(Ax - By, [A,B] \big).
\end{equation}

\section{Observer-independent Quantities}
In this appendix we show how one obtains observer-independent quantities (e.g. mass and spin) from the Lie algebra of the symmetry group. The presentation here is heavily influenced from  \cite{Simms}.

Let $G$ be a connected Lie group representing the symmetries on a spacetime manifold $M$. The two main examples to keep in mind is the Poincar{\'e} group, $\R^4 \rtimes \text{O}(1,3)$, and the Milne-like symmetry group, $\R \times \text{O}(1,3)$.

 The connection to physics is through the Lie algebra $\mathfrak{g}$ of $G$. Suppose $R\colon G \to \rm{U}(\mc{H})$ is a unitary representation. Here $\mathcal{H}$ is a separable Hilbert space associated with our spacetime manifold $M$, e.g. a space of `nice' functions on $M$ equipped with a suitable $L^2$ norm. For each $X \in \mathfrak{g}$, the one parameter subgroup $\R \to G$, $t \mapsto \exp (tX)$, is mapped into a strongly continuous one parameter subgroup $\R \to \rm{U}(\mc{H})$, $t \mapsto R \circ \exp(tX)$. By Stone's Theorem \cite[Theorem 10.15]{Hall}, there is a unique densely defined self-adjoint operator
 \begin{equation}\label{def of self-adjoint R}
 \dot{R}_X\colon \text{Dom}(\dot{R}_X) \to \mc{H} \:\:\:\: \text{ such that } \:\:\:\: R \circ \exp(tX) = \exp (-it\dot{R}_X)
 \end{equation}
  where the exponential on the right is defined by the spectral theorem for unbounded self-adjoint operators. The dense domain $\text{Dom}(\dot{R}_X)$ is the set of all $\psi \in \mathcal{H}$ such that the limit
\begin{equation}
\dot{R}_X\psi = i\lim_{t \to 0}\frac{R \circ \exp(tX)\psi - \psi}{t}
\end{equation}
exists in the norm topology. Since self-adjoint operators represent physical observables, each element $X \in \mathfrak{g}$ is associated with a physical quantity.

 An element $g \in G$ induces a \emph{change of observer} determined by the action of $g$ on $M$, $g \colon M \to M$. The transformation $ghg^{-1} \colon M \to M$ appears the same to the new observer as the transformation $h\colon M \to M$ did to the old observer. The transformation $h \mapsto ghg^{-1}$ is an inner automorphism $I_g$ of $G$, and so yields a homomorphism $I \colon G \to \rm{Aut}(G)$, $g \mapsto I_g$. Physically $I_g$ is the transformation of $G$ induced by a change $g$ of observer. The automorphism $I_g$ induces a Lie algebra automorphism
\begin{equation}
\text{Ad}_g\colon \mathfrak{g} \to \mathfrak{g}
\end{equation}
where $\mathfrak{g} = \text{Lie}(G)$ and $\rm{Ad}_g$ is the derivative of $I_g$ at the identity $e \in G$, i.e. $\text{Ad}_g = (dI_g)_e\colon \mathfrak{g} \to \mathfrak{g}$. This yields a group homomorphism 
\begin{equation}
\text{Ad}\colon G \to \text{Aut}(\mf{g}), \:\: g\mapsto \text{Ad}_g
\end{equation}
called the \emph{adjoint representation} of $G$.

The physical interpretation is that $\text{Ad}_gX$ is the physical quantity which appears the same to the new observer as $X$ did to the old observer. Therefore if $R \colon G \to \rm{U}(\mc{H})$ is a unitary representation, the self-adjoint operator $\dot{R}_{\text{Ad}_gX}$ represents the same quantity to the new observer as $\dot{R}_X$ does to the old where $\dot{R}$ is given by (\ref{def of self-adjoint R}). 

The goal is to find elements $X \in \mathfrak{g}$ such that $\text{Ad}_g(X) = X$ for all $g \in G$. This would imply that $X$ yields a physical quantity which does not depend on the observer. However there could also be invariant quantities which arise as polynomials in $\dot{R}_X$, and we would like to find these too. Therefore we want to expand the domain of $\text{Ad}_g$. The natural setting for this is the universal enveloping algebra.

The \emph{universal enveloping algebra} of $\mf{g}$ is the quotient
\begin{equation}
\mc{U}(\mathfrak{g}) = \mc{T}(\mf{g})/\mc{J}(\mf{g})
\end{equation}
where $\mc{T}(\mf{g})$ is the \emph{tensor algebra} of $\mathfrak{g}$ 
\begin{equation}
\mc{T}(\mf{g}) = \bigoplus_{n = 0}^\infty \mf{g}^{\otimes n} = \R \oplus \mf{g} \oplus (\mf{g} \otimes \mf{g}) \oplus (\mf{g} \otimes \mf{g} \otimes \mf{g}) \oplus \dotsb
\end{equation}
and $\mc{J}(\mf{g})$ is the two-sided ideal 
\begin{equation}
\mc{J}(\mf{g}) = \text{span}\big\{T_1 \otimes \big(X \otimes Y - Y \otimes X - [X,Y]\big) \otimes T_2 \mid X,Y \in \mf{g}, \:\: T_1,T_2 \in \mc{T}(\mf{g})\big\}.
\end{equation}
$\mathcal{U}(\mf{g})$ has the following universal property: any linear map $\phi \colon \mathfrak{g} \to A$ from $\mf{g}$ to a unital associative algebra $A$ satisfying $\phi\big([X,Y]\big) = \phi(X)\phi(Y) - \phi(Y)\phi(X)$ extends to a unique homomorphism $\tilde{\phi}\colon \mc{U}(\mf{g}) \to A$. For elements in $\mf{g}^{\otimes n}$ , the extension is given by $\tilde{\phi}(X_1 \otimes \dotsb \otimes X_n)= \phi(X_1)\dotsb \phi(X_n)$ and linear over each direct sum. 
%

Let $R\colon G \to \rm{U}(\mc{H})$ be a unitary representation and consider $\dot{R}$ from (\ref{def of self-adjoint R}). From \cite[p.20]{Simms} and references therein, there is a dense set $D_R \subset \mc{H}$ on which $\dot{R}_X$ is essentially self-adjoint for all $X \in \mathfrak{g}$; hence $\dot{R}_X$ is determined by its restriction to $D_R$. Let $\text{S}(D_R)$ be the Lie algebra of symmetric operators with domain $D_R$. Then 
\begin{equation}
\dot{R}\colon \mathfrak{g} \to \text{S}(D_R), \:\:\:\: X \mapsto \dot{R}_X
\end{equation} 
is a Lie algebra homomorphism. Therefore, by the universal property for $\mc{U}(\mf{g})$, $\dot{R}$ extends to a unique homomorphism 
\begin{equation}
\dot{R}\colon \mc{U}(\mf{g}) \to \text{Op}(D_R)
\end{equation}
where $\text{Op}(D_R)$ is the unital associative algebra generated by elements in $\text{S}(D_R)$, i.e. it's sums and products of essentially self-adjoint operators $\dot{R}_X$, $X \in \mf{g}$. 

Let $g \in G$ be a change of observer. The automorphism $\text{Ad}_g\colon \mathfrak{g} \to \mf{g}$ extends to a unique automorphism $\text{Ad}_g\colon \mc{U}(\mf{g}) \to \mc{U}(\mf{g})$. Let $U \in \mc{U}(\mf{g})$. Like above, the physical interpretation is that $\text{Ad}_gU$ is the physical quantitiy which appears the same to the new observer as $U$ did to the old observer. If $\dot{R}_U$ is the operator associated with the old observer, then $\dot{R}_{\text{Ad}_gU}$ is the operator associated with the new observer.

An element $C \in \mathcal{U}(\mf{g})$ is \emph{observer-independent} (also known as an \emph{invariant of $G$}) if $\text{Ad}_gC= C$ for all $g \in G$. In this case the eigenvalues of $\dot{R}_C$ yield \emph{observer-independent quantities}. The physical significance of $\dot{R}_C$ is independent of the choice of observer $g \in G$. From \cite[p. 33]{Simms} we have

\medskip

\begin{thm}\label{observer ind quantities}
Let $G$ be a connected Lie group. The observer-independent elements coincides with the center of $\mc{U}(\mf{g})$. 
\end{thm} 

\medskip

In general elements $C$ in the center of a universal enveloping algebra of a Lie algebra $\mf{g}$ are called \emph{Casimir elements}. A necessary and sufficient condition for $C\in \mc{U}(\mf{g})$ to be a Casimir element is
\begin{equation}
CX = XC, \:\:\:\: \text{ for all } X \in \mathfrak{g} / \mc{J}(\mf{g}) \approx \mf{g}.
\end{equation}
If $R\colon G \to \rm{U}(\mc{H})$ is a unitary representation, then $\dot{R}_C$ is called a \emph{Casimir operator}. If $R$ is irreducible, then $\dot{R}_C$ is a constant multiple of the identity by Schur's lemma \cite[Theorem 3.5 (a)] {Folland}.

\subsection{Casimir Elements for $\R^4 \rtimes \rm{SL}(2,\C)$} \label{casimir for poincare}

 The Lie algebra for the Poincar{\'e} group is $\R^4\oplus \mf{so}(1,3)$ with Lie bracket given by (\ref{lie algebra for poincare}). We choose the canonical basis $\{\pi_0, \pi_1, \pi_2, \pi_3\}$ for $\R^4$ and $\{m_{01}, m_{02}, m_{03}, m_{12}, m_{13}, m_{23}\}$ as a basis for $\mf{so}(1,3)$ where
\begin{equation}\label{boost basis for o(1,3)}
m_{01}=\left(\begin{array}{cccc} 
0 & 1 & 0 & 0
\\
1 & 0 & 0 & 0
\\
0 & 0 & 0 & 0
\\
0 & 0 & 0 & 0
\end{array}\right),
 \:\:\:\:\:
m_{02} = \left(\begin{array}{cccc} 
0 & 0 & 1 & 0
\\
0 & 0 & 0 & 0
\\
1 & 0 & 0 & 0
\\
0 & 0 & 0 & 0
\end{array}\right),
 \:\:\:\:\:
m_{03}  \left(\begin{array}{cccc} 
0 & 0 & 0 & 1
\\
0 & 0 & 0 & 0
\\
0 & 0 & 0 & 0
\\
1 & 0 & 0 & 0
\end{array}\right).
 \:\:\:\:\:
 \end{equation}
 
 \begin{equation}\label{rotation basis for o(1,3)}
 m_{12} = \left(\begin{array}{cccc} 
0 & 0 & 0 & 0
\\
0 & 0 & 1 & 0
\\
0 & -1 & 0 & 0
\\
0 & 0 & 0 & 0
\end{array}\right),
 \:\:\:\:\:
 m_{13} = \left(\begin{array}{cccc} 
0 & 0 & 0 & 0
\\
0 & 0 & 0 & 1
\\
0 & 0 & 0 & 0
\\
0 & -1 & 0 & 0
\end{array}\right),
 \:\:\:\:\:
 m_{23} = \left(\begin{array}{cccc} 
0 & 0 & 0 & 0
\\
0 & 0 & 0 & 0
\\
0 & 0 & 0 & 1
\\
0 & 0 & -1 & 0
\end{array}\right).
\end{equation}
We also define $m_{i0} = m_{0i}$ and $-m_{ji} =m_{ij}$. Put $k_i = m_{0i}$ and $j_i = -\e^{ijk}m_{jk}$ where $\e^{ijk}$ is completely antisymmetric with $\e^{123} = 1$. Then $k_i$ are mapped into Lorentz boosts via the exponential map and $j_i$ are mapped into counterclockwise rotations in the $\pi_i\pi_j$-plane. We simply write $\pi_\mu$ for $(\pi_\mu, 0) \in \R^4 \oplus \mf{so}(1,3)$ and likewise $m_{\mu\nu}$ for $(0,m_{\mu\nu}) \in \R^4 \oplus \mf{so}(1,3)$. We can identify $m_{\mu\nu}$ with $\pi_\mu \otimes \pi_\nu - \pi_\nu\otimes \pi_\mu$ via the Minkowski metric $\eta$
\begin{equation}
m_{\mu\nu}X = \eta(\pi_\nu, X)\pi_\mu - \eta(\pi_\mu, X)\pi_\nu\:\:\:\: \text{ for } X \in \R^4.
\end{equation}
Then using (\ref{lie algebra for poincare}),   the Lie algebra relations are
\begin{align}
[m_{\mu\nu}, \pi_\a] &= m_{\mu\nu}\pi_\a = \eta_{\nu\a}\pi_\mu - \eta_{\mu\a}\pi_\nu \label{lie alg 1}
\\
[\pi_\a, \pi_\b] &= 0
\\
[k_i,k_j] &= -\e^{ijk}j_k
\\
[j_i,j_j] &= \e^{ijk}j_k
\\
[j_i,k_j] &= \e^{ijk}k_k.
\end{align}
Using (\ref{lie alg 1}), a calculation shows $\eta^{\mu\nu}(\pi_\mu\otimes \pi_\nu)$ commutes with each $m_{\mu\nu}$ and clearly it commutes with each $\pi_\mu$. Therefore $\eta^{\mu\nu}(\pi_\mu\otimes \pi_\nu)$ is a Casimir element. Also $\eta^{\mu\nu}(w_\mu \otimes w_\nu)$ is a Casimir element where $w_\mu = \frac{1}{2}\e^{\mu\nu\a\b}\pi_\nu \otimes m_{\a\b}$ is the \emph{Pauli-Lubanski} element, but this is harder to show.


Let $R \colon \R^4 \rtimes \text{SL}(2,\C) \to \text{U}(\mc{H})$ be a unitary representation. (\ref{def of self-adjoint R}) yields self-adjoint operators $P^\mu := \dot{R}_{\pi_\mu}$. We call $P^0$ the \emph{energy} operator and $P^i$ the \emph{linear momentum}  operators. Similarly $J^i := \dot{R}_{j_i}$ and $K^i := \dot{R}_{k_i}$ are the \emph{angular momentum} and \emph{boost} operators, respectively. The Casimir operators 
\begin{equation}
M^2 := -\dot{R}_{\eta^{\mu\nu}(\pi_\mu\otimes\pi_\nu)} = -P^\mu P_\mu \:\:\:\: \text{ and } \:\:\:\: W^\mu W_\mu= \dot{R}_{\eta^{\mu\nu}(w_\mu\otimes w_\nu)} 
\end{equation}
 correspond to observer-independent quantities. $M^2$ is called the \emph{mass-squared} operator.

\medskip

\subsection{Casimir Elements for $\R \times \rm{SL}(2,\C)$} \label{casimir for milne-like}
The Lie algebra for $\R \times \text{SL}(2,\C)$ is $\R \oplus \mathfrak{so}(1,3)$ with Lie bracket 
\begin{equation}\label{milne lie bracket}
\big[(\tau_1,A), (\tau_2,B)\big] = \big(0, [A,B]\big).
\end{equation}
We choose the canonical basis $\{\pi\}$ for $\R$ and $\{m_{01}, m_{02}, m_{03}, m_{12}, m_{13}, m_{23}\}$ as a basis for $\mathfrak{so}(1,3)$ given by 
(\ref{boost basis for o(1,3)}) and (\ref{rotation basis for o(1,3)}). Again we put $k_i = m_{0i}$ and $j_i = -\e^{ijk}m_{jk}$ which are mapped into Lorentz boosts and rotations via the exponential map. We simply write $\pi$ for $(\pi,0) \in \R \oplus \mf{so}(1,3)$ and $m_{\mu\nu}$ for $(0,m_{\mu\nu}) \in \R \oplus \mf{so}(1,3)$. The Lie algebra relations are 
\begin{align}
[m_{\mu\nu}, \pi] &= 0 \label{milne lie alg 1}
\\
[\pi, \pi] &= 0 \label{milne lie alg 2}
\\
[k_i,k_j] &= -\e^{ijk}j_k
\\
[j_i,j_j] &= \e^{ijk}j_k
\\
[j_i,k_j] &= \e^{ijk}k_k.
\end{align}
Equations (\ref{milne lie alg 1}) and (\ref{milne lie alg 2}) show $\pi$ is a Casimir element. In \cite{Barg2} Bargmann found the Casimir elements for $\text{SL}(2,\C)$ which will be Casimir elements for $\R \oplus \mathfrak{so}(1,3)$ since they will commute with $\pi$. They are $\delta^{ij}(j_i\otimes j_j - k_i\otimes k_j)$ and $\delta^{ij}(j_i\otimes k_j)$.

Let $R\colon \R \times \text{SL}(2,\C) \to \text{U}(\mc{H})$ be a unitary representation. Then (\ref{def of self-adjoint R}) yields self-adjoint operators $M := \dot{R}_{\pi}$, which we call the \emph{mass} operator, and $J^i := \dot{R}_{j_i}$ and $K^i := \dot{R}_{k_i}$ are the \emph{angular momentum} and \emph{boost} operators. The Casimir operators are \cite{Barg2}
\begin{align}
M &:= \dot{R}_{\pi},
\\
 Q &:= \dot{R}_{\delta^{ij}(j_i\otimes j_j - k_i\otimes k_j)} = \delta_{ij}(J^iJ^j - K^iK^j),
 \\
  S &:= \dot{R}_{\delta^{ij}j_ik_j} = \delta_{ij}J^iK^j.
\end{align} 
These correspond to observer-independent quantities.


\bibliographystyle{amsplain}

\end{document}